\documentclass[aps,pra,superscriptaddress]{revtex4}

\usepackage{amsfonts,amsmath,graphicx}
\usepackage{color}
\usepackage[T1]{fontenc}
\newcommand{\be}{\begin{equation}}
\newcommand{\ee}{\end{equation}}

\begin{document}


\title{High-entropy magnetism of murunskite}


\author{D. Tolj}
\affiliation{William H. Miller III Department of Physics and Astronomy, Johns Hopkins University, Baltimore, USA}
\affiliation{Laboratory for Quantum Magnetism, \'Ecole Polytechnique F\'ed\'erale de Lausanne, CH-1015 Lausanne, Switzerland}
\author{P. Reddy}
\affiliation{Department of Physics, Faculty of Science, University of Zagreb, Croatia}
\author{I. \v Zivkovi\'c}
\affiliation{Laboratory for Quantum Magnetism, \'Ecole Polytechnique F\'ed\'erale de Lausanne, CH-1015 Lausanne, Switzerland}
\author{L. Ak\v samovi\'c}
\affiliation{Institute of Solid State Physics, TU Wien, 1040 Vienna, Austria}
\author{J. R. Soh}
\affiliation{Laboratory for Quantum Magnetism, \'Ecole Polytechnique F\'ed\'erale de Lausanne, CH-1015 Lausanne, Switzerland}
\affiliation{Quantum Innovation Centre (Q.InC), Agency for Science Technology and Research, Singapore 138634}
\author{K. Kom\c edera}
\affiliation{AGH University of Krakow, Faculty of Physics and Applied Computer Science, 30-059 Kraków, Poland}
\affiliation{Mössbauer Spectroscopy Laboratory, Institute of Technology, University of the National Education Commission, 30-084 Kraków, Poland}
\author{I. Bia\l{}o}
\affiliation{Physik-Institut, Universität Zürich, CH-8057 Zürich, Switzerland.}
\author{C. M. N. Kumar}
\affiliation{Institute of Nuclear Physics PAN, 31-342 Kraków, Poland}
\author{T. Iv\v si\'c}
\affiliation{Department of Physical Chemistry, Ru\dj er Bo\v skovi\'c Institute, 10000 Zagreb, Croatia}
\author{M. Novak}
\affiliation{Department of Physics, Faculty of Science, University of Zagreb, Croatia}
\author{O. Zaharko}
\affiliation{Paul Scherrer Institut, CH-5232 Viligen, Switzerland}
\author{C. Ritter}
\affiliation{Institut Laue-Langevin, 38000 Grenoble, France}
\author{T. La Grange}
\affiliation{Laboratory for Ultrafast Microscopy and Electron Scattering, \'Ecole Polytechnique F\'ed\'erale de Lausanne, CH-1015 Lausanne, Switzerland}
\author{W. Tabi\'s}
\affiliation{AGH University of Krakow, Faculty of Physics and Applied Computer Science, 30-059 Kraków, Poland}
\author{I. Batisti\'c}
\affiliation{Department of Physics, Faculty of Science, University of Zagreb, Croatia}
\author{L. Forr\'o}
\email{lforro@nd.edu}
\affiliation{Stavropoulos Center for Complex Quantum Matter, University of Notre Dame, USA}
\author{H. M. R\o nnow}
\email{henrik.ronnow@epfl.ch}
\affiliation{Laboratory for Quantum Magnetism, \'Ecole Polytechnique F\'ed\'erale de Lausanne, CH-1015 Lausanne, Switzerland}
\author{D. K. Sunko}
\email{dks@phy.hr}
\affiliation{Department of Physics, Faculty of Science, University of Zagreb, Croatia}
\author{N. Bari\v si\'c}
\email{nbarisic@phy.hr}
\affiliation{Department of Physics, Faculty of Science, University of Zagreb, Croatia}
\affiliation{Institute of Solid State Physics, TU Wien, 1040 Vienna, Austria}


\begin{abstract}
Murunskite (K$_2$FeCu$_3$S$_4$) is a bridging compound between the only two known families of high-temperature superconductors. It is a semiconductor like the parent compounds of cuprates, yet isostructural to metallic iron-pnictides. Moreover, like both families, it has an antiferromagnetic (AF)-like response with an ordered phase occurring below $\approx 100$~K. Through comprehensive neutron, M\"ossbauer, and XPS measurements on single crystals, we unveil AF with a nearly commensurate quarter-zone wave vector. Intriguingly, the only identifiable magnetic atoms, iron, are randomly distributed over one-quarter of available crystallographic sites in 2D planes, while the remaining sites are occupied by closed-shell copper. Notably, any interpretation in terms of a spin-density wave is challenging, in contrast to the metallic iron-pnictides where Fermi-surface nesting can occur. Our findings align with a disordered-alloy picture featuring magnetic interactions up to second neighbors. Moreover, in the paramagnetic state, iron ions are either in Fe$^{3+}$ or Fe$^{2+}$  oxidation states, associated with two distinct paramagnetic sites identified by M\"ossbauer spectroscopy. Upon decreasing the temperature below the appearance of magnetic interactions, these two signals merge completely into a third, implying an orbital transition. It completes the cascade of (local) transitions that transform iron atoms from fully orbitally and magnetically disordered to homogeneously ordered in inverse space, but still randomly distributed in real space.
\end{abstract}

\pacs{}

\maketitle

\section{Introduction}

The question of emergent order is one of the overarching preoccupations of contemporary science~\cite{Schrodinger92,Anderson72}. This stunning emergency often arises from the interplay of various physical forces, giving rise to complex patterns and hierarchical structures. By delving into the origins of order from disorder, scientists strive to unlock the underlying principles governing the behaviour and properties of complex materials and systems.

Modern functional materials, including the two classes of high-temperature superconducting compounds, cuprates and pnictides, present an intriguing combination of physical and chemical electronic properties~\cite{Shen08,Yoshida12,Si16}. Chemical binding closes orbitals, while physical functionality depends on their being open. The end points of the ensuing continuum of possibilities are rock salt and alkali metals. In between, insulators are sought for their magnetic and optical properties, while conductors are investigated in the hope of circumventing dissipation, either by superconductivity or by topologically restricted surface states.

The functionality of magnetic insulators typically depends on the open $d$ orbitals of transition metal ions. A classic example is the ferrimagnetism of magnetite, Fe$_2$O$_3$, which also exhibits the ubiquitous complication that not all iron atoms are in the same oxidation (spin) state. In this work, we describe the antiferromagnetic-like order of murunskite (K$_2$FeCu$_3$S$_4$), a quasi-2D semiconductor of relatively simple structure. Nevertheless, it challenges the generic paradigm of magnetism in insulators, which relies on a direct connection between crystal structure and the location of magnetic moments.

In the wider setting of functional materials, murunskite interpolates between the high-temperature superconducting cuprates and pnictides~\cite{Tolj21}. The key distinction is in the role of the ligands. In cuprates, there is a charge transfer between copper $3d$ orbitals and the $2p$ orbitals of the ligand oxygens~\cite{Zaanen85,Eskes93}, which bridge neighboring coppers so that their $3d$ orbitals do not overlap. This charge transfer appears as a localized hole shared within this ``CuO$_4$ molecule,'' which plays a key role for understanding of cuprate normal and superconducting states~\cite{Barisic15,Pelc19,Barisic22,Kumar23}. In pnictides, the size of the unit cell is set by arsenic ligand orbitals bound with Fe $e_{g}$ suborbitals that are closed (ionic) and far from the Fermi level, while the metallicity~\cite{Barisic10,Kumar23} stems from the direct overlap of correlated $t_{2g}$ suborbitals on neighboring Fe ions~\cite{Eschrig09,Fink09}. Thus, in cuprates there is a clean separation between Fermi-liquid (conducting) and non-Fermi liquid (localized) charges, while in pnictides the metallicity is directly affected by the spin correlations stemming from the Hubbard repulsion in overlapping $t_{2g}$ suborbitals~\cite{Borisenko16,Derondeau17,Barisic22,Sunko20a}. These interactions in the spin channel may yet turn out to be responsible for SC in the pnictides, while the experimental evidence in the cuprates clearly points to the charge channel via the localized hole~\cite{Barisic15,Pelc19,Barisic22,Kumar23}. In murunskite, the metal sites are bridged by ligand S $2p$ orbitals. These are partially open even in the insulating parent compound, making it electronically more similar to the cuprates, even though it is isostructural with the pnictides~\cite{Tolj21}.

 In this article, a nearly commensurate magnetic order is established at $97$~K although the positions of the iron atoms in the crystal lattice are disordered. Thus, the magnetism of murunskite depends on the question of emergent order to quite an unusual, even critical extent, making it an interesting playground for the development of next-generation paradigms. In particular, real-space disorder cannot be relegated to an afterthought, but must be considered on an equal footing from the outset. In this respect, murunskite is reminiscent of high-entropy alloys~\cite{Moniri23}, although the crystal lattice of our samples is perfect, not deformed or glass-like at the structural level, within the error of the high-resolution TEM measurements. The problem we face is converse to that of high-entropy alloys: How can the randomness in the distribution of both kinds of magnetically active ions, Fe$^{2+}$ and Fe$^{3+}$, among the closed-shell Cu$^+$ \emph{not} affect either the structural or the magnetic order?

To address this question, we recapitulate already published results on the electronic and crystal structure of murunskite first, based on XPS and magnetic susceptibility measurements. Next, we present new results of neutron diffraction and M\"ossbauer spectroscopy. Finally, we discuss a possible resolution of the conundrum outlined above, based on relatively long-range magnetic interactions, up to second-neighbor, among the iron atoms~\footnote{We count neighbors by the number of orthogonal hops in our simulation, cf.\ Fig.~\ref{figsimulations}a.}. It presumes an active role of the sulfur ligands, which, we believe, will play a part in any complete solution of this problem. In conclusion, we argue that the isolation of local multi-centric wave functions as emergent building blocks of functional materials is a promising paradigm for future applications.

\section{Materials and methods}

Single crystals of murunskite were grown by two step synthesis as previously reported~\cite{Tolj21}. First, iron copper sulfide was prepared from elements by solid state reaction. Afterwards, elemental potassium was added, and single crystals were grown from the melt by slow cooling.

Powder neutron diffraction was measured on the D20 beamline at the Institut Laue Langevin in Grenoble (ILL). Approximately 1.1 g of powder was loaded in a vanadium container and measured with a wavelength of 2.4 \AA. A very high intensity 2-axis diffractometer equipped with a large microstrip detector at extremely high neutron flux allowed us to measure full powder diffraction at a base temperature, and the temperature dependence in an interesting temperature range.

A large ($6\times 7\times 4$~mm$^3$) murunskite single crystal was measured in neutron diffraction with a 4-circle mode on ZEBRA instrument at the Paul Scherrer Institut in Villigen (PSI). The short (1.18\AA) and long (2.32\AA) wavelength, with the two available detectors (point detector and 160mm x 160mm area detector) allowed a reliable determination of $k$-vectors.

The symmetry analysis was performed using ISODISTORT from the ISOTROPY software \footnote{H. T. Stokes, D. M. Hatch, and B. J. Campbell, ISOTROPY Software Suite, iso.byu.edu (2021).}\cite{Campbell06} and software tools of the Bilbao crystallographic server \cite{Aroyo11, Perez15}. The Rietveld refinement of neutron powder diffraction data to determine the crystal and magnetic structure parameters was performed using the Mag2pol program~\cite{Qureshi19}, with the use of its internal tables for neutron scattering lengths.

M\"ossbauer spectra were measured at the M\"ossbauer setup in Krakow, Poland to determine the local Fe environment and murunskite magnetic structure. M\"ossbauer spectra have been collected in standard transmission geometry for the $14.41$~keV transition in $^{57}$Fe using a commercial $^{57}$Co(Rh) source kept under ambient pressure and at room temperature. The absorbing sample was prepared in a form of powder by mixing 30 mg of murunskite powder with the B$_{4}$C carrier. The absorber thickness amounted to 14.9 mg/cm$^{2}$, the latter having natural isotopic composition. A Janis Research Co. SVT-400 cryostat was used to maintain the absorber temperature, with a long-time accuracy better than 0.01 K (except for 4.2 K, where the accuracy was better than 0.1 K). A RENON MsAa-4 M\"ossbauer spectrometer equipped with a Kr-filled proportional counter was used to collect spectra in the photo-peak window. The velocity scale of the M\"ossbauer spectrometer was calibrated by a Michelson-Morley interferometer equipped with the He-Ne laser. The spectral shifts are given in respect to the shift of natural $\alpha$-Fe at ambient pressure and room temperature. The spectra were fitted within transmission integral approximation using the Mosgraf-2009 software~\cite{Duraj11}.

\section{Known electronic and structural properties}

\begin{figure*}
\includegraphics[width=179mm]{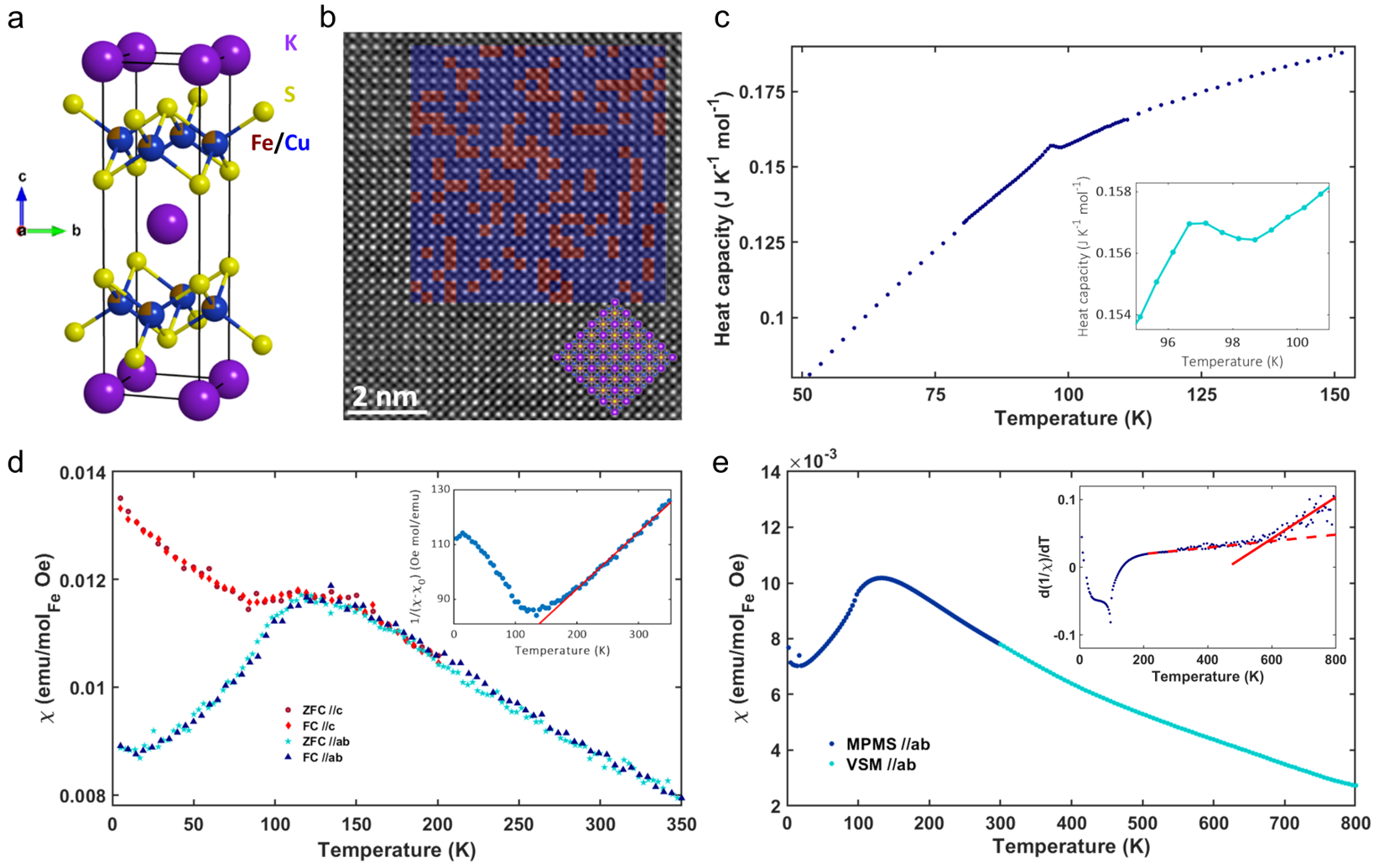}
\caption{(a) Murunskite crystal structure. Fe and Cu atoms share the same crystallographic position. (b) Atomic resolution TEM image of murunskite taken with a high-angle annular dark field (HAADF) detector. The overlay at top right shows a simulated random Fe/Cu distribution (red/blue, respectively) in a single layer, cf.\ Fig.~\ref{figsimulations}c. At bottom right is a 5x5 unit-cell overlay parallel to the c direction. (c) Heat capacity versus temperature, with a pronounced peak around the AFM transition, magnified in the inset. (d) Temperature dependence of the magnetic susceptibility with H = 1 T applied (FC) and zero field (ZFC) runs. The inset shows fitted result using the Curie–Weiss law, where the red line is the fitting curve for ZFC, H || ab measurement. (e) Temperature dependence of the magnetic susceptibility with H = 1 T applied for H || ab measured with MPMS (dark blue, T<300K) and VSM (light blue, T>300K). The inset shows the temperature dependence of the derivative of the magnetic susceptibility. (Red lines are guides to the eye, with full line indicating the change in slope above 550K).}
\label{Fig1}
\end{figure*}

Murunskite crystallizes in the tetragonal ThCr$_{2}$Si$_{2}$ structure type (space group I4/mmm, no. 139) with the lattice parameters a = b = 3.868(1) \AA, c = 13.079(9) \AA
(Fig.~\ref{Fig1}a)~\cite{Tolj21}. As grown crystals of murunskite (K$_2$FeCu$_3$S$_4$) were analyzed with Energy Dispersive X-ray Spectroscopy couple with Scanning Electron Microscopy (SEM-EDS) and Transmission Electron Microscopy (TEM-EDS). No ordering or clustering of iron was observed. Elemental analysis of the two areas with a most pronounced difference in iron content observed in TEM-EDX elemental maps, resulted in a variations on the range of 1 at\% which is within measurement error and expected range for random distribution. Elemental analysis can be found in Supplementary Information (Fig. S1 and Table S1).

The temperature dependence of the magnetic susceptibility and heat capacity are presented in Fig.~\ref{Fig1}~\cite{Tolj21}. The paramagnetic response at high temperature shows a Curie-Weiss-like behavior. Upon cooling below 150 K, this behavior is followed by a broad maximum which is characteristic for the appearance of strong short-range antiferromagnetic-like correlations in low dimensional, frustrated or disordered systems. Upon cooling below 100 K, long-range anisotropic ordering is observed with a splitting of in-plane (IP) and out-of-plane (OOP) curves. A transition temperature of 97 K was confirmed by the heat capacity measurement where the mild peak indicates a magnetic transition. No difference was observed between the field-cooled (FC) and zero-field-cooled (ZFC) protocols, showing no ferromagnetic contribution in the system.

Assuming that the magnetic moments reside exclusively on the Fe ions, a moment of $6.3$~$\mu_B$ \emph{per} atom was inferred by fitting to the AF Curie-Weiss law in the high-temperature window $170$--$300$~K. It is somewhat higher than the maximum calculated (spin-only) value ($5.92$~$\mu_B$) for Fe$^{3+}$, and much higher than the similar value for Fe$^{2+}$ ($4.89$~$\mu_B$). The magnitude of the fitted moment appears even more excessive when one considers the XPS data on the same crystals~\cite{Tolj21}, which show that at least\footnote{The cited measurements may have overestimated the Fe$^{3+}$ contribution due to surface oxidation.} $2/3$ of the Fe ions are in the $2+$ state. The same measurement clearly shows that copper is in the $1+$ (nonmagnetic $3d^{10}$) state. Additionally, high temperature data show no deviation from paramagnetic behavior up to 550 K, as can be seen in Fig.~\ref{Fig1} (e), in further support of the notion that the observed large moments are local.

In antiferromagnetic quasi-2D systems, such as murunskite, splitting of in-plane (IP) and out-of-plane (OOP) curves suggests that magnetic moments are oriented primarily within the ab plane~\cite{Blundell01}. However, the out-of-plane susceptibility continues to rise below the magnetic transition, while in-plane measurements fail to converge towards the zero at the base temperature, indicating a complex behavior of the system. 

\section{Neutron scattering}
The random distribution of magnetic irons in the $3d^{10}$ copper matrix raises the interesting question: where and how do spins organize in such a system? To this end, and to shed more light on the microscopic nature of magnetic anomalies observed in the magnetometry data presented earlier, murunskite was studied using powder and single-crystal neutron diffraction experiments.

\begin{figure*}[htb!]
\includegraphics[width=179mm]{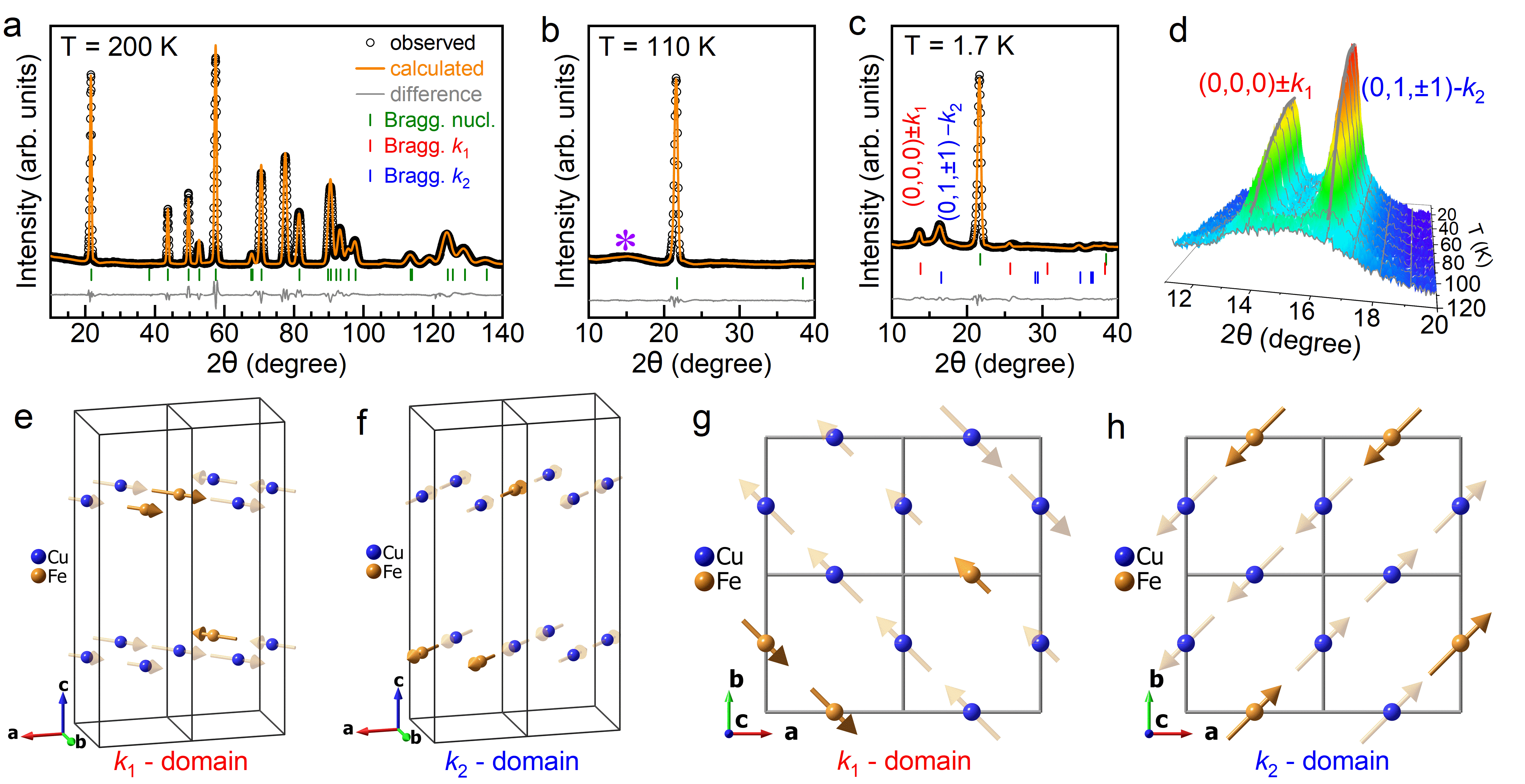}
\caption{{Neutron powder diffraction: NPD data at three representative temperatures 200~K (a), 110~K (b) and 1.7~K (c) along with Rietveld refinement results. At 200~K the observed data could be refined using only crystal lattice contribution. All diffraction peaks correspond to the expected Bragg positions for the lattice with the space group $I4/mmm$. The diffraction data and the refinements at 110~K are similar to the latter case, except for a broad diffuse scattering peak centered at $2\theta \approx 15^{\circ}$ due to short-range magnetic order. (c) The refinement of 1.7~K data was carried out by assuming two equally populated magnetic domains with propagation vector $k_1$ or $k_2$, with best MSSG described in the main text. (d) A contour plot of the low-angle portion of NPD data showing the temperature evolution of diffuse scattering peak and two main magnetic peaks associated with the two magnetic domains with propagation vector $k_1$ or $k_2$. Graphical representation of spin arrangements obtained from the Rietveld refinements, in two magnetic domains within 2x1x1 crystallographic unit cells (e--f) and within 2x2x0.5 unit cells (g--h).}
}
\label{fig_neutron_data}
\end{figure*}

The neutron diffraction data and Rietveld refinement results for three distinct temperatures are presented in Fig.~\ref{fig_neutron_data}(a-c). The NPD data in the paramagnetic regime was refined with lattice-only contribution (Fig.~\ref{fig_neutron_data}(a)). An onset of short-range correlations was observed at $150$~K, characterized by a broad diffuse scattering peak centered around $2\theta = 15^{\circ}$ (see Fig.\ref{fig_NPD_Supplement}(c) in Appendix~\ref{Append:NPD_SymmAna}). In Fig.~\ref{fig_neutron_data}(b) a portion of the NPD data and refinements at 110~K highlights the short-range order diffuse scattering peak. With the decrease in temperature from 110~K, the intensity of the diffuse scattering peak reduces and several new Bragg peaks were observed below $100$~K due to a long-range magnetic order. Two new Bragg peaks appear on top of the diffuse scattering peak. As the temperature is decreased the intensity of these new Bragg peaks grows at the expense of the diffuse scattering peak until they saturate around $40$~K. Fig.~\ref{fig_neutron_data}(d) shows the temperature-dependent behavior of NPD data surrounding the diffuse scattering peak in the temperature range of 110 - 1.7~K. This captures all three main features associated with the evolution of magnetism with temperature namely, the short-range diffuse scattering peak, the onset of long-range order with two main magnetic peaks below 100~K, and the saturation of the magnetic peaks around 40~K. It is worth noting that the magnetic susceptibility data in Fig.~\ref{Fig1}(d) begins to deviate from the Curie-Weiss law at a temperature of approximately $150$~K, which notably coincides with the onset of the diffuse scattering peak in NPD at 150~K.

Neutron diffraction on a single crystal in 4-circle mode enabled us to find and index the Bragg peaks in the parent lattice setting I4/mmm. Two magnetic propagation vectors were identified to be close to the quarter zone vectors  $k_1$= (0.25, 0.25, 0)  (k-point (a a 0) and $k_2$ = (0.25, 0.75, 0) (k-point (a 1-a 0). The profile fitting procedure of neutron powder diffraction data using the $k$-vectors obtained from single crystals did not match the observed pattern, but a good fit was obtained assuming two close but different incommensurate propagation vectors $k_1$=(0.266(3),0.266(3),0) and $k_2$=(0.24(2),0.76(2),0)=(0.24(2),1-0.24(2),0). Magnetic structure was then resolved by using these two $k$-vectors that can be attributed to either a single magnetic domain with two propagation vectors or two magnetic domains with a single propagation vector. As is well known, neutron diffraction alone cannot determine whether the observed multiple $k$-vectors are due to two magnetic domains, or form a single domain with a multi-$k$-vector structure. As a consequence, the refinements in both cases will produce similar results. Here we use a two-domain, single-$k$ approach to resolve the magnetic structure, assuming an equal population of $k_1$ and $k_2$ magnetic domains. The details of the symmetry analysis and all possible magnetic superspace groups (MSSG) associated with both magnetic domains are presented in the Appendix~\ref{Append:NPD_SymmAna}.

To determine the IC magnetic structures, we used the mag2pol~\cite{Qureshi19} for a detailed Rietveld refinement of the NPD diffraction pattern against four possible maximal magnetic superspace groups (MSSG) for $k_1$ domain and two possible MSSGs for the $k_2$ domain described in the Appendix~\ref{Append:NPD_SymmAna}. We obtained best fits to the experimental data with MSSGs Fmmm.1'(0,0,g)s00s (irrep - mDT3, mk7t4) and B2/m.1'(a,b,0)00s (irrep - mC1, mk2t1) for the $k_1$ and $k_2$ magnetic domains, respectively. The results of the refinements are presented in Fig.~\ref{fig_neutron_data}(c) and the resulting magnetic models are presented in Fig.~\ref{fig_neutron_data}(e-h). In both MSSGs, the out-of-plane (OOP) magnetic moments are allowed. The refinements indicated negligible c-components, which were then fixed to be zero, thus the moments lie within the $ab$-plane. In the $k_1$ domain, the nearest-neighbor spins along the $c$ direction are arranged in parallel, while they are antiparallel in the $k_2$ domain. From the refinements, the total amplitude of magnetic moment per Fe$^{2+}$ was found to be $3.02(7)\mu_{\rm{B}}$ and $3.06(3)\mu_{\rm{B}}$ in $k_1$ and $k_2$ domains, respectively at 1.7~K. This indicates that the moments have similar total amplitude in both domains despite the differences in magnetic modulations.

Remarkably, the faded arrows in Fig.~\ref{fig_neutron_data}(e-h) do not, in fact, refer to anything physical. The refinement determines the average magnetic moments per Wyckoff site without differentiating Cu from Fe. Sample stoichiometry, TEM measurements, single crystal XRD, and high-temperature neutron data all concur that $3/4$ of the 4d Wyckoff sites are randomly occupied by Cu$^+$ ions in the nonmagnetic $3d^{10}$ state. Furthermore, the Fe ions are predominantly in the $2+$ state, while the remainder are in the $3+$ state, and they are all scattered at random throughout the lattice. The central question for the magnetism of murunskite, parallel to that in related materials such as iron-pnictides, is where the physical magnetic moments reside, and are they local (orbital) moments or spin-density waves. In order to shed further light on this issue, we turn to M\"ossbauer spectra.

\section{M\"ossbauer spectra}

\begin{figure*}
\begin{tabular}{cc}
\includegraphics[width=175mm]{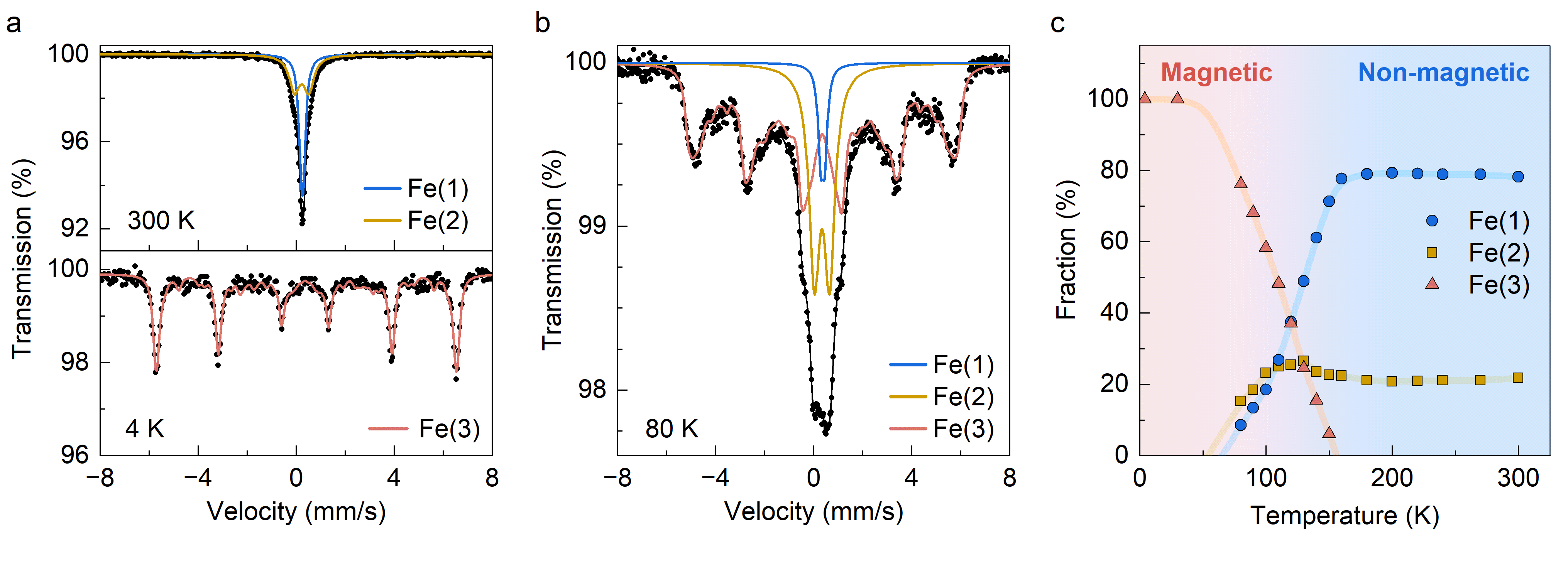}
\end{tabular}
\caption{Selected $^{57}$Fe M\"ossbauer spectra at three representative temperatures (a) 300K and 4K, (b) 80K. The experimental data are represented by black dots, and the solid lines represent the fitting to the experimental data. Blue and yellow lines corresponds to the two paramagnetic Fe(1) and Fe(2) sites, respectively, and red represents magnetic Fe(3) site. See the text and SM for details and complete dataset. (c) Temperature evolution of the contribution from each of the nonequivalent Fe sites. The colors of the solid points correspond to the lines in the fitting depicted in panels (a) and (b). The shaded lines are guides to the eye.}
\label{moessbauer_data}
\end{figure*}

Fitted M\"ossbauer spectra at few characteristic temperatures together with temperature dependence of observed iron sites fraction are presented in Fig.~\ref{moessbauer_data}. The overall temperature dependence of M\"ossbauer data agrees well with magnetization and heat capacity measurements. It additionally reveals two paramagnetic sites above 150 K and one magnetic site appearing under that temperature in the murunskite powder sample. 

The fractions of these two paramagnetic sites are roughly 80:20, and they are nearly temperature independent until the third, magnetic site appears at about 150 K, which we associate with onset of short-range ordering. The larger of the two paramagnetic fractions decreases steadily below this onset. The smaller one remains the same until the magnetic transition to long-range order at $97$~K, after which both decrease steadily. On the other hand, the magnetic site exhibits a monotonic increase in fraction level from the point at which it first appears until it reaches 100\% around $30$~K, indicating again that all iron sites become equivalent and apparently fully ordered magnetically at low temperatures, despite the random distribution of local moments with two different valence states. This observation is in good agreement with the neutron powder measurement that show magnetic peak intensities saturating around 40K. The uniqueness of the magnetic site would seem to mitigate in favor of a multi-$k$ vector over a multidomain structure. However, the M\"ossbauer dipole response is not really sensitive to the microscopic origin of the long-range field, while a multidomain structure appears naturally in the simulations below, so we argue in favor of the latter in the Discussion.

The random Fe/Cu distribution provides the setting for the various environments. A simple statistical analysis shows that the proportion of irons on a quadratic lattice which have only copper as closest neighbors is 28\%, while there are 6\% of those with $3$ or $4$ Fe neighbors. The fraction of Fe ions in the $3+$~state as observed in XPS is at most $1/3$. The simplest hypothesis is based on counting electrons in different Fe environments in a sulfide layer. Fe ions surrounded by only Cu nearest-neighbors account for the Fe$^{3+}$ state. Fe ions with 4 Fe nearest-neighbors would push iron below the Fe$^{2+}$ oxidation state and are not preferred as found in the reported end compound of the series $\approx$K$_{0.8}$Fe$_{1.6}$S$_{2}$, where the structure is only stabilised by partial occupation of K and Fe positions~\cite{Lei11-180503}. That leaves Fe ions with mixed nearest-neighbor ions to be in the $2+$~state.

The next clue comes from the observation that the dominant paramagnetic (atomic) site~$1$ in M\"ossbauer spectroscopy is sensitive to the onset of short-range order, while the minority paramagnetic site $2$ is only sensitive to long-range order. Both are depleted in favor of the single magnetic site~$3$. We infer that Fe$^{2+}$ ions are responsible for the incipient ordering fluctuations at $150$~K, while the Fe$^{3+}$ ions begin to participate in the magnetic order only when it becomes long-range. It is an open question at present whether the saturation of the magnetic site at 100\% below $40$~K corresponds to a Fe$^{3+}\to$~Fe$^{2+}$ orbital transition, meaning that all magnetic sites belong to a single orbital state at very low temperature (for more details see Appendix~\ref{spectrappendix}).

\section{Simulations}

In disordered alloys, the Brillouin zone of the parent lattice can survive in angle-resolved photoemission spectroscopy long after alloying has nominally destroyed translational invariance~\cite{Popescu12}. The real-space disorder primarily affects the peak widths in inverse space, which  retain some intensity as long as the material is not structurally amorphous. We believe that we are observing a similar phenomenon here and documenting it for the first time in the magnetic sector. To substantiate this belief, we turn to simulations. Details are given in Appendix~\ref{simappendix}.

\begin{figure*}
\includegraphics[width=175mm]{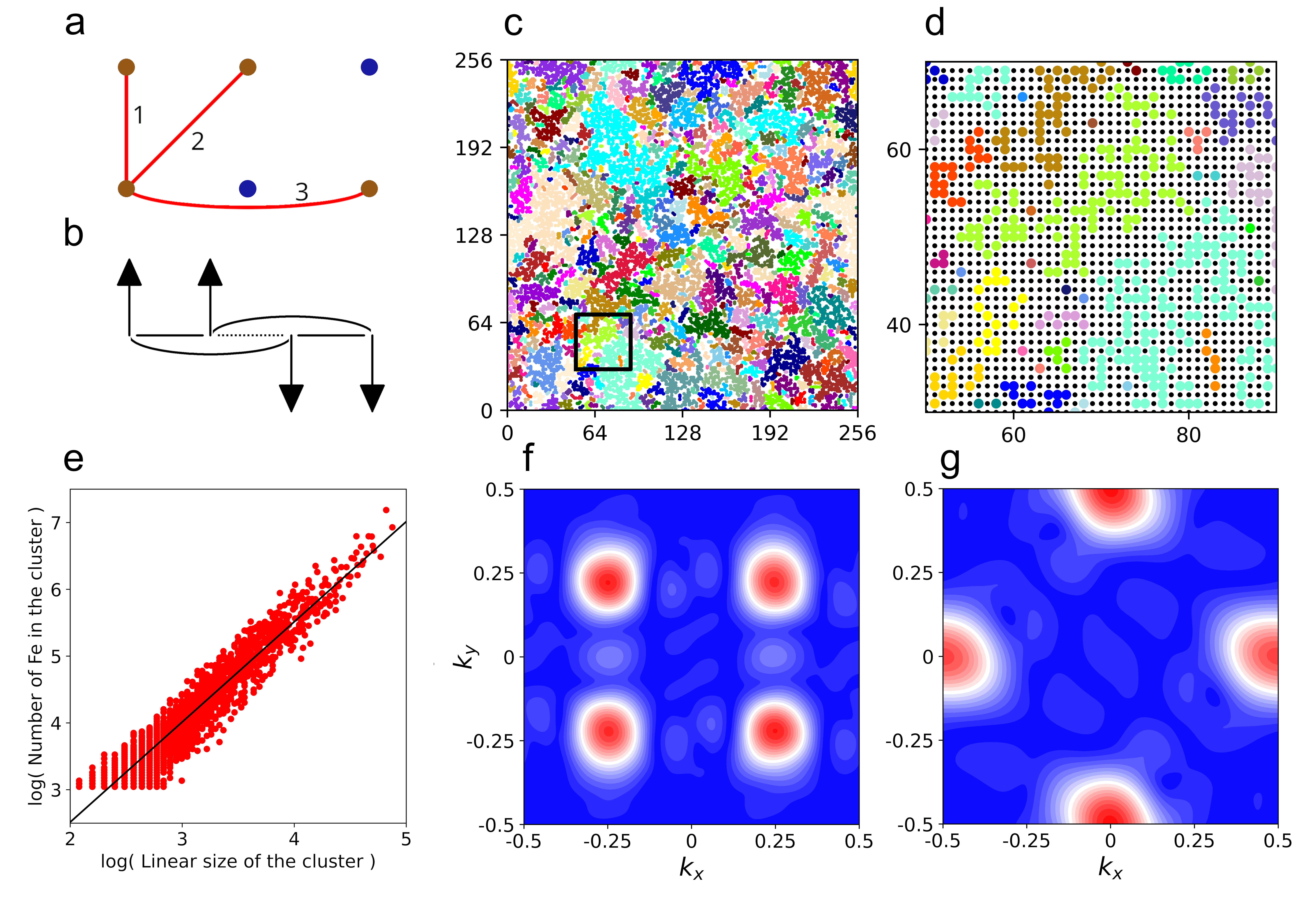}
\caption{(a) Spin-spin interactions attributed to sites both occupied by Fe atoms, where the first-neighbor interaction $J_1>0$ and both second-neighbor interactions $J_{2,3}<0$. (b) A quarter-zone configuration with two satisfied $J_1$ and $J_3$ bonds and one frustrated $J_1$ bond, with total energy $-J_1+2J_3$. (c) Magnetic clusters in a $256\times 256$-sample simulation near T$_c$. The Fourier transform of a similar $1024\times 1024$ pattern appears in panel (f). Distinct clusters of Fe ions percolating through any of the bonds in (a) are depicted in different colors. The empty spaces are non-magnetic Cu ions. The colored sites are $1/4$ of all sites, even if they dominate visually because the dots are sized to touch as nearest neighbors. Black square: inset magnified in (d) reveals that each percolating cluster is like a tree of cracks in the non-magnetic background of Cu ions, shown as black dots. (e) Scaling of the number of Fe atoms vs. linear size (side of enveloping square) of a cluster, showing a fractal dimension of $1.50$ for clusters with more than $20$~Fe atoms. The regression line is $(1.5001\pm 0.0092)x-0.4856\pm 0.0280$. (f) Fourier transform of a simulated spin distribution for $J_2=J_3=-1.5 J_1$. Compare the real-space pattern in panel (c). (g) Fourier transform of a distribution with $J_2=-1.5 J_1$ and $J_3=0$. Units of $2\pi/a$.}
\label{figsimulations}
\end{figure*}

Starting with a random distribution of Fe atoms, we attribute spin-spin interactions according to Fig.~\ref{figsimulations}a to all positions where any of the red lines in the figure connects two Fe atoms. We solve the ensuing Heisenberg Hamiltonian in the mean-field approximation on a lattice of $1024 \times 1024$ sites.

In order to narrow the possibilities for the coupling constants, we performed DFT calculations in large supercells. It turns out that $J_1$ has to be ferromagnetic (positive), leading to Fe atoms forming magnetic dimers when they are next to each other. These dimers each carry a rather large moment, roughly double the moment of the Fe$^{2+}$ ions, which explains the high magnetic moment observed in the magnetic susceptibility measurement. $J_2$ is found to be negative, while our supercells are not large enough to include $J_3$, so we cannot infer anything about it from DFT.

In Fig.~\ref{figsimulations}f, we have a clear indication that $J_3$ has to be of the same order and sign as $J_2$. With this assignment, our simulation reproduces neutron scattering satisfactorily. The precise positions of the (nearly) quarter-zone peaks depend on the values of the coupling constants rather weakly, say putting $J_2=-J_1$ moves them less than their width, about the same as varying the stoichiometry from FeCu$_3$ to Fe$_{1.1}$Cu$_{2.9}$, the latter being roughly what we have in the synthesized sample. Depending on the actual random distribution of the Fe sites with all other parameters equal, the $k$-space maxima showed a variation between $0.24$ and $0.27$ on either axis. Between this and the experimental uncertainty, we think that further attempts to improve the fits would bring diminishing returns.

On the other hand, a qualitatively different Fourier pattern appears when setting $J_3=0$, as shown in Fig.~\ref{figsimulations}g. Hence, the inference that a significant linear second-neighbor coupling is present in the material is quite robust. This result may be compared to ferropnictides' magnetic responses invariably being like in Fig.~\ref{figsimulations}g, i.e. half-zone rather than quarter-zone. In pnictides, all lattice sites are occupied by magnetic Fe atoms, so the magnetic subsystem is expectedly simpler.

The same simulation gives $T_c\approx 4.8J_1$, the prefactor weakly dependent on the actual distribution of Fe atoms. With the measured $T_c\approx 100$~K, this amounts to $J_1\approx 0.002$~eV. That value is significantly smaller than the superexchange in cuprates ($\sim 0.1$~eV)~\cite{Eskes93}, or antiferromagnetic pnictides ($\sim 0.05$~eV)~\cite{Zhao09}, subject to two cautions. First, our simulations do not include quantum fluctuations, which would require a larger $J_1$ for the same $T_c$. Second, a lower $J_1$ than in oxides is qualitatively expected from the lower energy differences and greater covalency of sulfur orbitals. We conclude that the effective $J_1$ is underestimated relative to the real one because of the limitations of the simulation. Indirect experimental evidence for that can be found in the high temperature area of the magnetic susceptibility shown in Fig.~\ref{Fig1}e. The inset shows a slope change in susceptibility data above $\sim 550$ K, which can be attributed to dissociation of Fe-Fe dimers. This temperature corresponds to a microscopic $J_1\sim 0.05$~eV, which is in the same range as in the antiferromagnetic pnictides.

Previous DFT calculations have shown that the S orbitals in murunskite are partially open even in the insulating state, with a small spin polarization~\cite{Tolj21}. Such orbitals are a natural conduit for magnetic correlations, which, as we see, must extend at least to both second neighbors to account for the data.

\section{Discussion and conclusions}

The magnetic and by extension orbital order of murunskite emerges in a crystal whose structural lattice order coexists with total disorder in the magnetic mixed-valence Fe ion positions, which occupy $1/4$ of the available tetrahedral sites, the remaining $3/4$ being taken by magnetically inert Cu$^+$ ions in the closed-shell $3d^{10}$ configuration. Taking that substrate into account, the magnetic responses are surprisingly regular. Calorimetry, neutron scattering and M\"ossbauer spectroscopy concur that there is an AF transition at $97$~K with a single ordered site at low temperatures.

We have proposed a scenario in which these properties coexist naturally. The main statistical idea is that a lot of real-space disorder can be subsumed into peak widths in inverse space, so murunskite is like a disordered alloy magnetically, even if it shows none of the local distortions and incipient glassiness in the crystal structure \cite{Tolj21}, characteristic of high-entropy alloys. We can account for neutron scattering quantitatively within this simple and novel framework.

M\"ossbauer and XPS data present a more nuanced picture. Only one ordered magnetic site emerges from two distinct paramagnetic ones at high temperature, which saturates at $100$\% at $40$~K, deep below the AF transition temperature. The two paramagnetic sites may be associated with Fe atoms, either in different environments, or different oxidation states, most probably a combination of the two. The key question is how the single ordered site emerges. We envisage two concurrent processes, one of them based on simulations, the other a conjecture subject to experimental verification.

As our simulations approach the transition temperature from above, we observe the emergence of magnetically ordered islands (percolating patches), each much smaller than the system as a whole, and separated by domain walls of isolated sites, as shown in Fig.~\ref{figsimulations}d. Notably, if we add a third-neighbor interaction $J_4$ in the simulation, a single patch of the size of the system appears already at T$_c$, which is difficult to reconcile with M\"ossbauer data, so we also have an observational indication that $J_4\approx 0$. If $J_4$ is very small, these islands may begin coupling below T$_c$, to cover the whole system below $40$~K, in accord with the data showing a gradual depletion of the paramagnetic sites below T$_c$. In addition, smaller clusters have a smaller T$_c$, which also spreads out the crossover region between $150$ and $40$~K. In cuprates, the fluctuation region also has a percolative character \cite{Pelc19,Barisic15}. However, because the Fe positions are quenched, the cluster sizes are fixed here, so the fluctuation range above T$_c$ is very wide. Nevertheless, the thermodynamic AF transition at T$_c=97$~K is unambiguous from a confluence of calorimetric, neutron, and M\"ossbauer (site $2$) data.

Because the patches are AF, and incoherent among themselves, they amplify the neutron signal as if we had a number of small samples, perfectly aligned crystallographically. The two wave-vectors $k_1$=(0.266(3),0.266(3),0) and $k_2$=(0.24(2),1-0.24(2),0) determined from powder diffraction data are similar, so it is most natural to interpret the AF state as a multidomain structure with some phase slips between the coherent patches.

The second process is a gradual orbital transition (crossover) from Fe$^{3+}$ to Fe$^{2+}$ between $97$ and $40$~K. In this scenario, only the Fe$^{2+}$ ions participate in the long-range order, as indicated by the depletion of the majority paramagnetic site beginning already at 150 K. This scenario is subject to experimental verification, which is not easy with XPS at low temperature because of sample charging. A similar orbital-transition phenomenon is the key for understanding the cuprates~\cite{Barisic15,Pelc19,Kumar23,Barisic22}. There, a crossover from $p$ to $1 + p$ mobile charges is observed as the doping p changes from $p = 0.1$ to $p = 0.3$, because the localized hole delocalizes. The salient point in the present context is that the same transition is also observed as a function of temperature~\cite{Barisic15,Pelc19,Kumar23,Klebel23}, over a range of $100$--$200$~K, so it is not unreasonable that an orbital crossover should occur in murunskite as well, only more difficult to observe directly because the material is insulating. In that case, the 100\% saturation of the M\"ossbauer signal below $40$~K would be naturally explained.

The magnetism of murunskite has turned out to be different than that of pnictides or parent-compound cuprates, owing in the first place to the real-space disorder of the magnetic Fe ions. It raises the question of the microscopic coherence of the magnetic percolating clusters. In hole-doped cuprates, superexchange-mediated AF order in the parent compound is rapidly degraded by hole doping associated with local disorder, so it could also be percolative near the end of its range. The appearance of ferromagnetically coupled dimers of Fe$^{2+}$ ions at nearest-neighbor sites has been a robust feature of our DFT calculations. It indicates that multi-centric wave functions involving the partially open S-ligand $2p$ orbitals are the basic building blocks of the clusters. The participation of polarized ligand orbitals in the total magnetic moment pushes murunskite outside of the usual paradigm of superexchange-mediated AF insulators in the direction of molecular magnets, in accord with the greater covalency of sulfur with respect to oxygen bonding.

The AF of the small tree-like patches (islands) discovered here is extremely robust, because it develops intrinsically against total underlying disorder, which is an interesting property in the context of material design for technological applications. As shown in Fig.~\ref{figsingle_clusters} in Appendix~\ref{simappendix}, as soon as a simulated cluster is large enough to have a Fourier response, it is similar to the whole sample. Thus, the small island size opens intriguing possibilities for (meta)material fabrication and manipulation. First is nanotexturing, controlling the size of the islands by impurities or substrate effects. Second is the realization that each island is a fractal quantum dot with potentially interesting electronic properties in its own right. Third, the material is easy to cleave. All three observations naturally point to 2D realizations in thin films, even single-layer, buttressed by the recent realization that low dimensionality is not such an obstacle to ordering as commonly assumed~\cite{Palle21}.

To conclude, we observe emergent long-range magnetic order in murunskite despite complete site and orbital disorder of the magnetic ions involved. We propose that it is a magnetic analogue of a disordered alloy. The microscopic foundation of this proposal is a functionalization of partially open S orbitals in mediating magnetic interactions between Fe ions. A mesoscopic cluster structure is inferred to evolve in the material deep below the AF transition temperature. A possible orbital transition of the Fe ions in parallel with this evolution remains to be confirmed by direct observation.

\pagebreak
\appendix
\section{Compositional analysis}

Energy Dispersive X-ray Spectroscopy coupled with Scanning Electron Microscopy (SEM-EDX) measurements were performed on as-grown single crystals that had been mechanically extracted from the ceramic bulk before being mounted on a SEM aluminium holder covered in carbon tape. No oxygen peak was observed in the EDS spectra. The composition acquired from SEM-EDX data served as the reference composition for data analysis on all subsequent measurements. The transmission Electron Microscopy sample was prepared by partially crushing as-grown single crystals in the agate mortar and placing the resulting flakes in high-purity isopropanol (argon blown for 20 minutes to decrease the amount of dissolved oxygen). After dispersing the flakes in the solution with an ultrasonicator, the sample was drop casted on Pure Silicon TEM Windows (Electron Microscopy Sciences) and allowed to dry overnight in an argon-filled glovebox. Differences in SEM and TEM element quantification are likely due to inaccuracy of an internal TEM calibration, sample preparation and increased measurement error for light elements.

\begin{figure}[!ht]
\includegraphics[width=150mm]{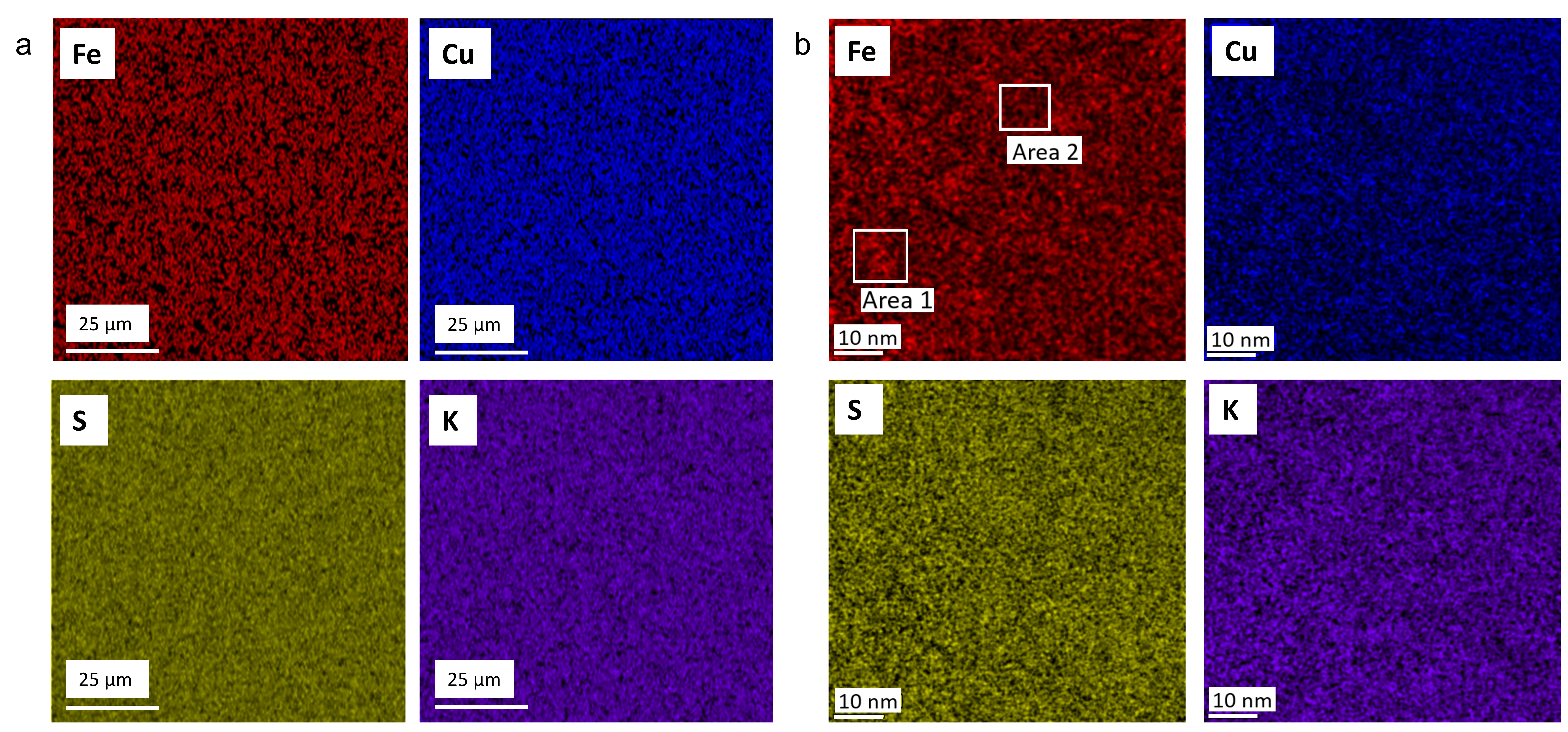}
\caption{Elemental maps  showing the random element distribution obtained by (a) SEM-EDS and (b)TEM-EDS with marked analysed areas on sample 1.}
\end{figure}

\begin{table}[!ht]
\includegraphics[width=80mm]{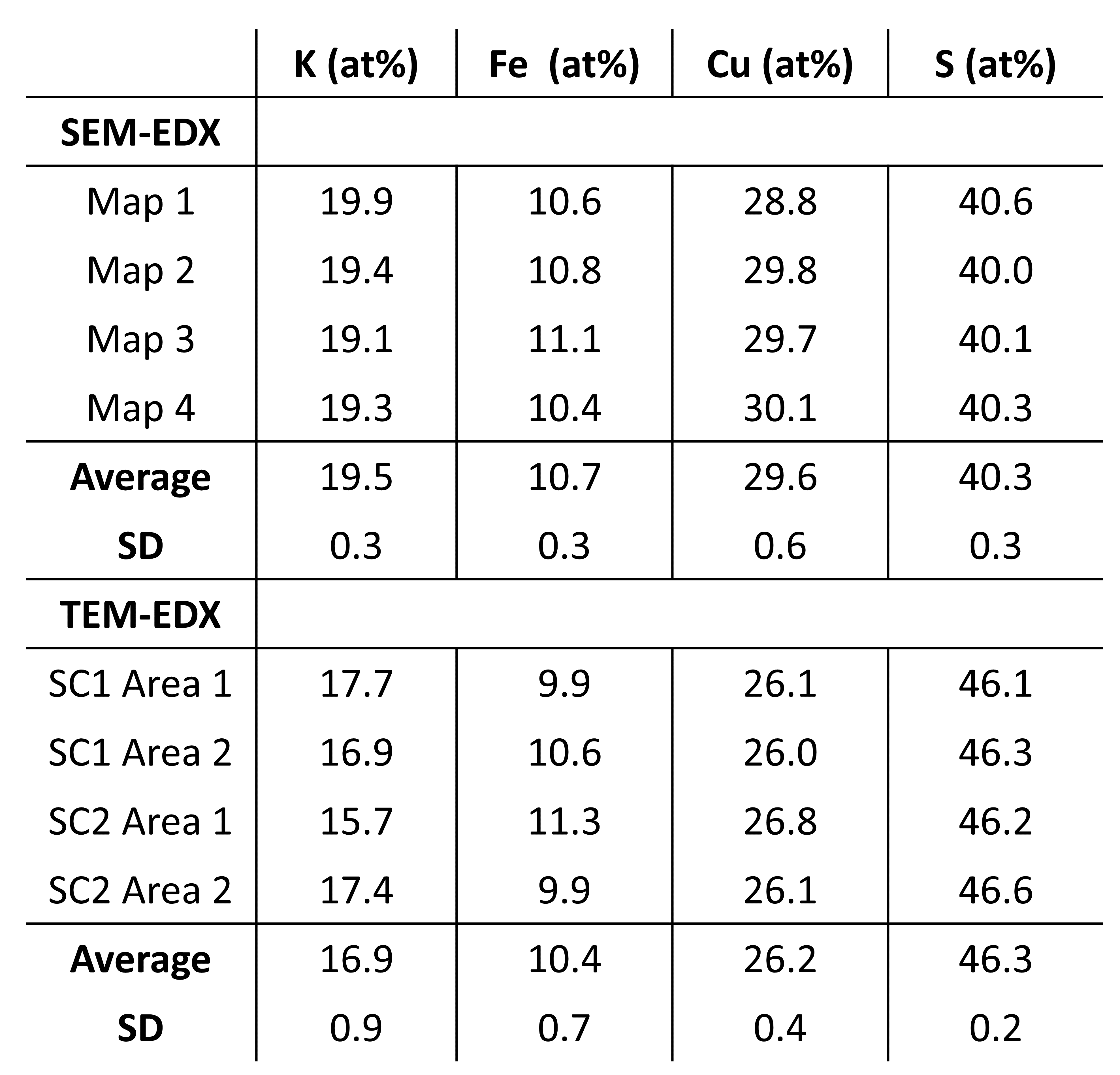}
\caption{Elemental quantification of (a) SEM-EDS elemental maps, and (b) TEM-EDS maps with marked areas for reported quantification.}
\end{table}
\pagebreak


{
\section{Neutron diffraction, symmetry analysis and magnetic models}
\label{Append:NPD_SymmAna}
}

\begin{figure*}[!ht]
\includegraphics[width=135mm]{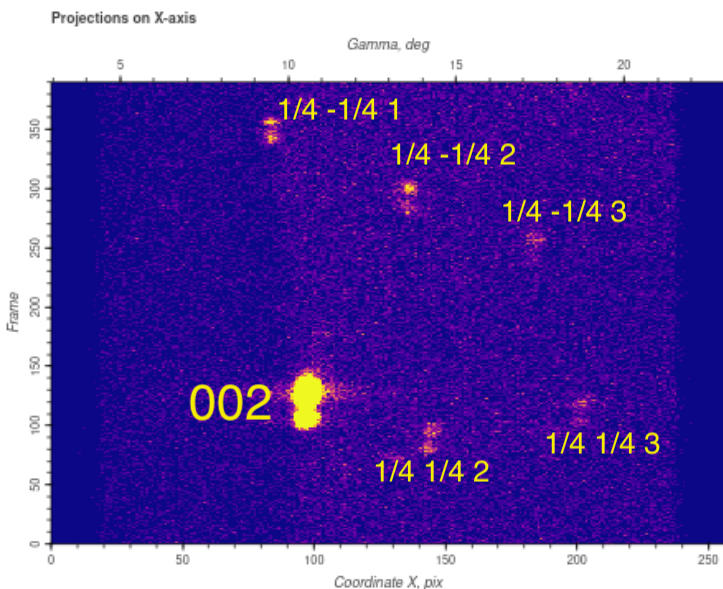}
\caption{Example of the reciprocal space map used to determine magnetic propagation vectors collected on ZEBRA beamline at PSI}
\label{NeutronZEBRA}
\end{figure*}

{The small number of magnetic peaks made it impossible to determine the magnetic structure reliably using only neutron powder diffraction (NPD). Figure~\ref{NeutronZEBRA} shows an example of the acquired reciprocal space map in single crystal neutron diffraction at base temperature, with magnetic and nuclear 002 reflections. The k-vectors were identified as approximately k1 = 0.25, 0.25, 0 and k2 = 0.25, 0.75, 0. The Bragg spots are spread due to the stacking faults in the layered materials and presence of mosaic grains in the measured crystal which precluded precise determination of $k$-vectors components. The profile fitting procedure of neutron powder diffraction data using the $k$-vectors obtained from single crystals did not match the observed pattern, but a good fit was obtained assuming two close but different incommensurate propagation vectors $k_1$=(0.266(3),0.266(3),0) and $k_2$=(0.24(2),1-0.24(2),0). The symmetry analysis and the refinements of magnetic structures using NPD data were subsequently done using these incommensurate $k$-vectors.}

\begin{figure*}[htb!]
\includegraphics[width=179mm]{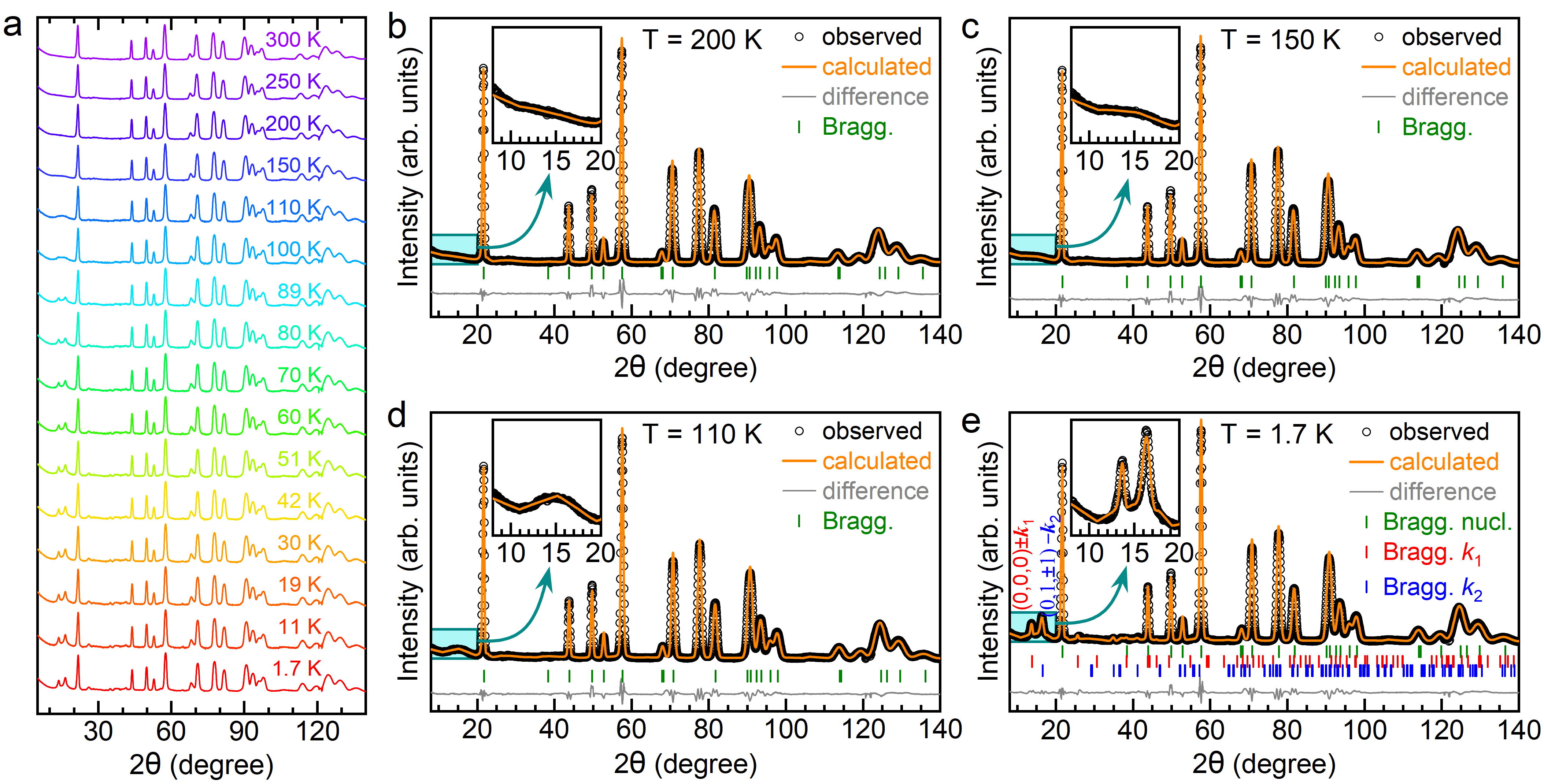}
\caption
{Neutron powder diffraction: (a) A waterfall plot of NPD data at selected temperatures, the intensities are shown in logarithmic scale for clarity. Rietveld refinement results of the NPD data at (b) $200$~K, (c) $150$~K (d) $110$~K and (e) $1.7$~K. At $200$~K the observed data could be refined using the crystal lattice contribution alone. The inset of panels (b)-(e) show the enlarged portion of NPD data where the most intense magnetic peaks are expected. The crystal structure was refined with the tetragonal space group $I4/mmm$, at all temperatures studied. A broad diffuse scattering peak centered at $2\theta \approx 15^{\circ}$ is observed at $150$~K which is further enhanced at $110$~K. The refinement of 1.7~K data was carried out assuming two equally populated magnetic domains with propagation vector $k_1$ or $k_2$.}

\label{fig_NPD_Supplement}
\end{figure*}

 To elucidate the temperature evolution of magnetic order, the NPD data were collected in the temperature range $1.7-300$~K, with denser temperature points below $110$~K ($1$~K intervals). A few selected data sets are presented in Fig.~\ref{fig_NPD_Supplement}(a), with the intensities shown in logarithmic scale, for clarity. From Fig.~\ref{fig_NPD_Supplement}(a) and (c) it is clear that we start to see a broad diffuse scattering peak centered around $2\theta \approx 15^{\circ}$ at $150$~K which can be attributed to a short-range ordering. With reduction in temperature the intensity of the diffuse scattering peak is further enhanced down to $110$~K (Fig.~\ref{fig_NPD_Supplement}(d)). The onset temperature of short-range magnetic order is surprisingly high compared to long-range magnetic order observed around $100$~K, but consistent with the M\"ossbauer spectroscopy.

By taking the $I4/mmm$ space group as the parent structure, we performed magnetic symmetry analyses for the domains with propagation vector $k_1 = (0.265, 0.265, 0)$ and $k_2 = (0.235, 0.765, 0)$, using ISODISTORT tool~\footnote{H. T. Stokes, D. M. Hatch, and B. J. Campbell, ISOTROPY Software Suite, iso.byu.edu (2021).}\cite{Campbell06}. For the case of $k_1$ there are four maximal MSSG of type (3 + 1) with 2 arms, and for $k_2$ there are two maximal MSSG of the type (3+1) with four arms, listed in table \ref{tab:MSSG}. Among four MSSGs of the magnetic domain with $k_1$, models Fmmm.1'(0,0,g)000s and Fmmm.1'(0,0,g)ss0s predict zero intensity for the most intense magnetic peak $(0, 0, 0)\pm k_1$ so can be are not suitable models. The refinements using model Fmmm.1'(0,0,g)s00s with the basis~\{(-1,1,0,0),(0,0,1,0),(1,1,0,0),(0,0,0,1)\} indicate that the magnetic moments are aligned nearly along c-axis $c$-axis with a small canting angle, this contradicts the magnetization results on single crystals. The remaining MSSG for $k_1$ domain is Fmmm.1'(0,0,g)s00s with the basis~\{(0,0,1,0),(-1,1,0,0),(-1,-1,0,0),(0,0,0,1)\}. Out of two MSSGs for $k_2$ domain, the model B2/m.1'(a,b,0)0ss leads to the moments nearly along the $c$-axis and thus can be ignored due to the reasoning described above. Thus we used Fmmm.1'(0,0,g)s00s with the basis~\{(0,0,1,0),(-1,1,0,0),(-1,-1,0,0),(0,0,0,1)\} and B2/m.1'(a,b,0)00s as magnetic models for $k_1$ and $k_2$ magnetic domains, respectively. These models resulted in the best fits to the experimental data in the temperature range $1.7-100$~K, and the refinement results of $1.7$~K data using the best magnetic models are presented in Fig.~\ref{fig_NPD_Supplement}(e). The best fitting MSSG models for both $k_1$ and $k_2$ domains allow magnetic moment along crystallographic $c$-axis, but the initial refinements indicated negligible magnetic components along $c$ so was fixed to be zero. This resulted in magnetic models with spins aligned within the crystallographic $ab$-plane along the diagonal (Fig.~\ref{fig_neutron_data}(e)-(h)). 

\begin{table}[htb!]
\centering
\caption{The irreps and MSSG of the type (3+1) obtained from the software ISODISTORT. The magnetic Fe occupies 1/4 of $4d$ wyckoff sites with coordinates [0, 0.5, 0.25] in the parent $I4/mmm$ crystal lattice with lattice parameters $a=b=3.8624(6)$~{\AA} and $c=13.0308(4)$~{\AA} at 1.7~K.}
\begin{tabular}{ccccc}
\hline
$k$-active			&	irrep		&	MSSG number			&	MSSG name			&	basis \\ \hline
$k_1$=(0.265,0.265,0)	&	mDT1, mk7t1	&	69.1.17.1.m522.2	&	Fmmm.1'(0,0,g)000s	&	\{(-1,1,0,0),(0,0,1,0),(1,1,0,0),(0,0,0,1)\} \\
					&	mDT2, mk7t3	&	69.1.17.2.m522.2	&	Fmmm.1'(0,0,g)s00s	&	\{(-1,1,0,0),(0,0,1,0),(1,1,0,0),(0,0,0,1)\} \\
					&	mDT3, mk7t4	&	69.1.17.2.m522.2	&	Fmmm.1'(0,0,g)s00s	&	\{(0,0,1,0),(-1,1,0,0),(-1,-1,0,0),(0,0,0,1)\} \\
					&	mDT4, mk7t2	&	69.1.17.3.m522.2	&	Fmmm.1'(0,0,g)ss0s	&	\{(-1,1,0,0),(0,0,1,0),(1,1,0,0),(0,0,0,1)\} \\ \hline
$k_2$=(0.235,0.765,0)	&	mC1, mk2t1	&	12.1.4.1.m59.2		&	B2/m.1'(a,b,0)00s	&	\{(1,-1,0,0),(0,-1,0,0),(0,0,-1,0),(0,0,0,1)\} \\
					&	mC2, mk2t2	&	12.1.4.2.m59.2		&	B2/m.1'(a,b,0)0ss	&	\{(1,-1,0,0),(0,-1,0,0),(0,0,-1,0),(0,0,0,1)\} \\ \hline
\end{tabular}
\label{tab:MSSG}
\end{table}

\section{M\"ossbauer spectroscopy\label{spectrappendix}}
\begin{figure*}[!ht]
\includegraphics[width=12cm]{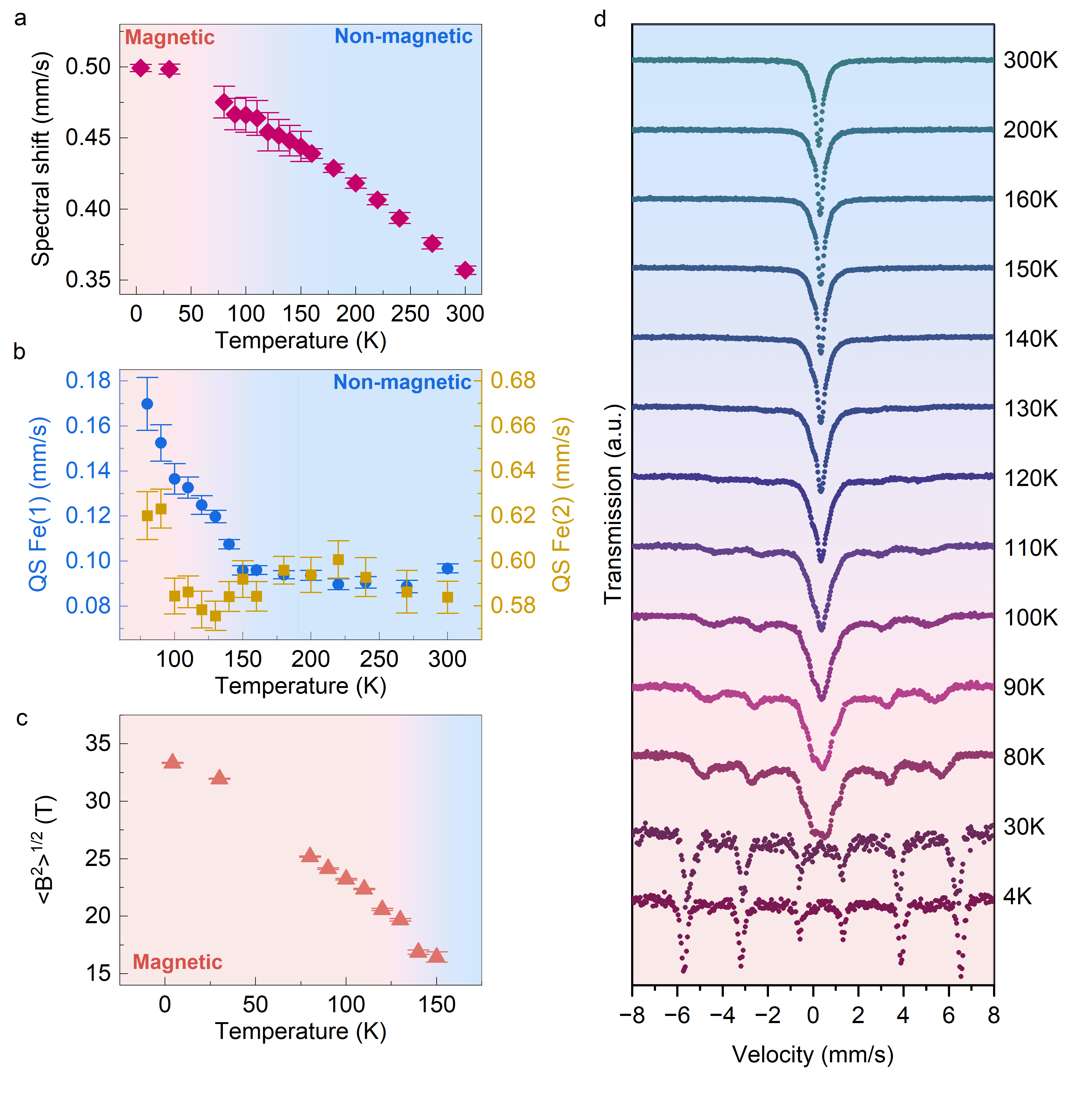}
\caption{The temperature dependence of the parameters extracted from  $^{57}$Fe M\"ossbauer spectroscopy; (a) spectral shift, (b) quadrupole splitting of Fe(1) and Fe(2) sites, and (c) mean squared amplitudes of the hyperfine magnetic field. Blue and yellow dots in (b) correspond to the two paramagnetic Fe(1) and Fe(2) sites, respectively. (d) Temperature evolution of the $^{57}$Fe M\"ossbauer spectra. Blue and red shadow regions in each of the plots correspond to non-magnetic and magnetic temperature regimes, respectively. }
\label{moessbauer_quadrupolar}
\end{figure*}
The two paramagnetic (atomic) sites may represent two distinct iron environments, but they cannot be directly attributed to different Fe valence states, although those states are identified by X-ray photoemission. Further information can be gathered from quadrupolar splittings and isomer shifts in Fig.~\ref{moessbauer_quadrupolar}. The isomer shift is directly related to s electron density at the nucleus and shielding by the d electrons. The shift for both sites is similar and smaller than what is expected for simple high spin Fe$^{2+}$ state confirming complex covalent interactions\cite{Gütlich2012}. With each quadrupolar splitting corresponding to the sites fraction behavior, we can see a smooth evolution of fitting parameters with temperature. The calculated Debye temperature is $458\pm 38$~K, which is typical for this material class. The mean value of the hyperfine field can be estimated with reasonable accuracy, and it is found to be $31.97$~T at a base temperature. By monitoring the change in magnetic moment distribution, we can conclude that it is widespread. The magnetic structure in real space cannot be resolved using M\"ossbauer data. By combining it with nuclear diffraction measurements, we infer murunskite's magnetic structure and exchange interactions.

There is only one sextet at lower temperatures, indicating a single ordered magnetic structure. To fit the data, a spatially modulated magnetic structure or a distribution of various hyperfine interactions are used, and a specific mean is used to approximate the characteristic sextet. A transmission integral approximation has been applied to fit M\"ossbauer spectra exhibiting magnetic hyperfine interaction. The magnetic modulated structure should be viewed as modulations of the electron spin polarization. The absorption profile of the magnetic modulated spectral component was processed by applying a quasi-continuous distribution of the magnetic hyperfine field B. For the Fe magnetic moment being collinear with the hyperfine field, the amplitude of the magnetic modulated structure along the direction parallel to the wave vector can be expressed as a series of odd harmonics:
\begin{equation}
B(qx)=\sum_{n=1}^N h_{2n-1}\sin[(2n-1)qx]
\end{equation}
where B(qx), q, x, and qx denote respectively the magnetic hyperfine field arising from magnetic modulated structure, the wave number of magnetic modulated structure, the relative position of Fe ion along the propagation direction of the stationary magnetic modulated structure, and phase shift. The symbols $h_{2n-1}$ denote the amplitudes of subsequent harmonics. The index N enumerates the maximum number of relevant harmonics. Amplitudes up to six subsequent harmonics (N = 6) have been fitted to the spectral shape. The constant number of N harmonics was chosen in such a way as to obtain a good-quality fit of the experimental spectrum represented by the parameter $\chi^2$ per degree of freedom of the order of 1.0. The argument $qx$ satisfies the following condition: $0\le qx\le 2\pi$ due to the periodicity of the magnetic modulated structure.

\section{Simulations\label{simappendix}}

Assume that antiferromagnetic (AF) ordering can be described by an extended isotropic Heisenberg model:
\begin{equation}
        H = -\sum\limits_{\langle ij \rangle} J_{ij}
                        \vec{S}_i \cdot \vec{S}_j
                - \sum\limits_{i} g_S \mu_B \vec{B} \cdot \vec{S}_i.
\end{equation}
In the mean-field approximation (MFA), the spin interactions are substituted by a site-dependent (local) magnetic field acting on each spin, $\vec{B}_i^{(loc)}$:
\begin{equation}
	H \approx -\sum\limits_{i} g_S \mu_B
	\underbrace{\left( \vec{B}_i^{(loc)} + \vec{B} \right)}_{\vec{B}_i^{(tot)} }
	\cdot \vec{S}_i + \text{ const.}
\end{equation}
The local magnetic field is generated by the average value of surrounding spins through the spin-spin interaction:
\begin{equation}
        g_S \mu_B \vec{B}_i^{(loc)} =  \sum\limits_j J_{ij} \langle {\vec{S}_j} \rangle.
\end{equation}
The average value of spin operators can be calculated from the total magnetic field acting on the spin by Curie's law:
\begin{equation} \label{eq:scf}
        \langle {\vec{S}_i} \rangle
        = \frac{ C }{ T } 
	\underbrace{
        \left( g_S \mu_B \vec{B} + \sum\limits_j J_{ij} \langle {\vec{S}_j} \rangle \right)
	}_{\sim \vec{B}_i^{(tot)} },
\end{equation}
where $C = S(S+1)/3k_B$. Eq. \eqref{eq:scf} is a system of equations in the unknown average values of site spins, $\langle {\vec{S}_i} \rangle$, for the given external magnetic field $\vec{B}$:
\begin{equation} \label{eq:4chi}
	\frac{ T }{ C }\langle {\vec{S}_i} \rangle
	-\sum\limits_j J_{ij} \langle {\vec{S}_j} \rangle
	= g_S \mu_B \vec{B}.
\end{equation}
In the high temperature limit, the spin average is different from zero only for a finite magnetic field. At low temperatures it is possible to have a non zero solution even if the magnetic field is zero. This spontaneous magnetization (non zero spin average) appears for temperatures below some critical temperature, $T_c$, which appears as the highest eigenvalue of the following problem:
\begin{eqnarray} \label{eq:eigen}
        \sum\limits_j J_{ij} \langle {\vec{S}_j}  \rangle = 
        \frac{T_c}{C} \cdot \langle {\vec{S}_i} \rangle.
\end{eqnarray}
The eigenvector corresponding to the highest eigenvalue describes the spin configuration at $T_c$.

For the cubic lattice with only nearest-neighbour interaction $J_{ij}$, and where this interaction is a positive constant, the largest eigenvalue corresponds to the eigenvector where all $\langle \vec{S}_i \rangle$ are equal. This eigenvector describes a ferromagnetic ordering. However, if the nearest-neighbour interaction is a negative constant, the largest eigenvalue corresponds to the eigenvector with alternating (positive/negative) values of $\langle \vec{S}_i \rangle$, describing an antiferromagnetic ordering.

The same approach can be applied to any system of interacting spins: ferromagnetic, antiferromagnetic, frustrated and with a random interaction $J_{ij}$. For a real system it implies finding the highest eigenvalue of a very large matrix $J_{ij}$.

Now we can turn to our problem: the magnetic ordering in the lattice where only 1/4 of sites are occupied by a spin (an iron ion) in an random way.

The problem was solved numerically. The spin configuration (distribution over lattice sites) is generated by the random number generator using \emph{python-numpy} package. The spin interaction is described by three constants, $J_1$, $J_2$ and $J_3$, as illustrated in Fig.~\ref{figsimulations}a. From DFT calculations it is known that nearest neighbor iron ions tend to align spin in a ferromagnetic order. Therefore we shall assume that $J_1$ is positive. Also, DFT calculations suggest that the next nearest neighbor iron ions tend to align spins antiferromagnetically. Consequently, we choose $J_2$ to be a negative constant.

For the given spin configuration, the matrix $J_{ij}$ is generated by using these three interaction terms. The lattice was a single 2D layer of $1024\times 1024$ sites, with periodic boundary conditions.

For the square lattice where only one quarter of Cu sites is randomly replaced by Fe, and with the finite-range interaction reduced to three nearest neighbour terms ($J_1$, $J_2$, $J_3$), the Fe ions create finite-size clusters of mutually interacting Fe ions, with sizes significantly smaller that the lattice size. That is illustrated in Fig.~\ref{figsimulations}c for a $256\times 256$ lattice.

\begin{figure*}
\includegraphics[height=7cm]{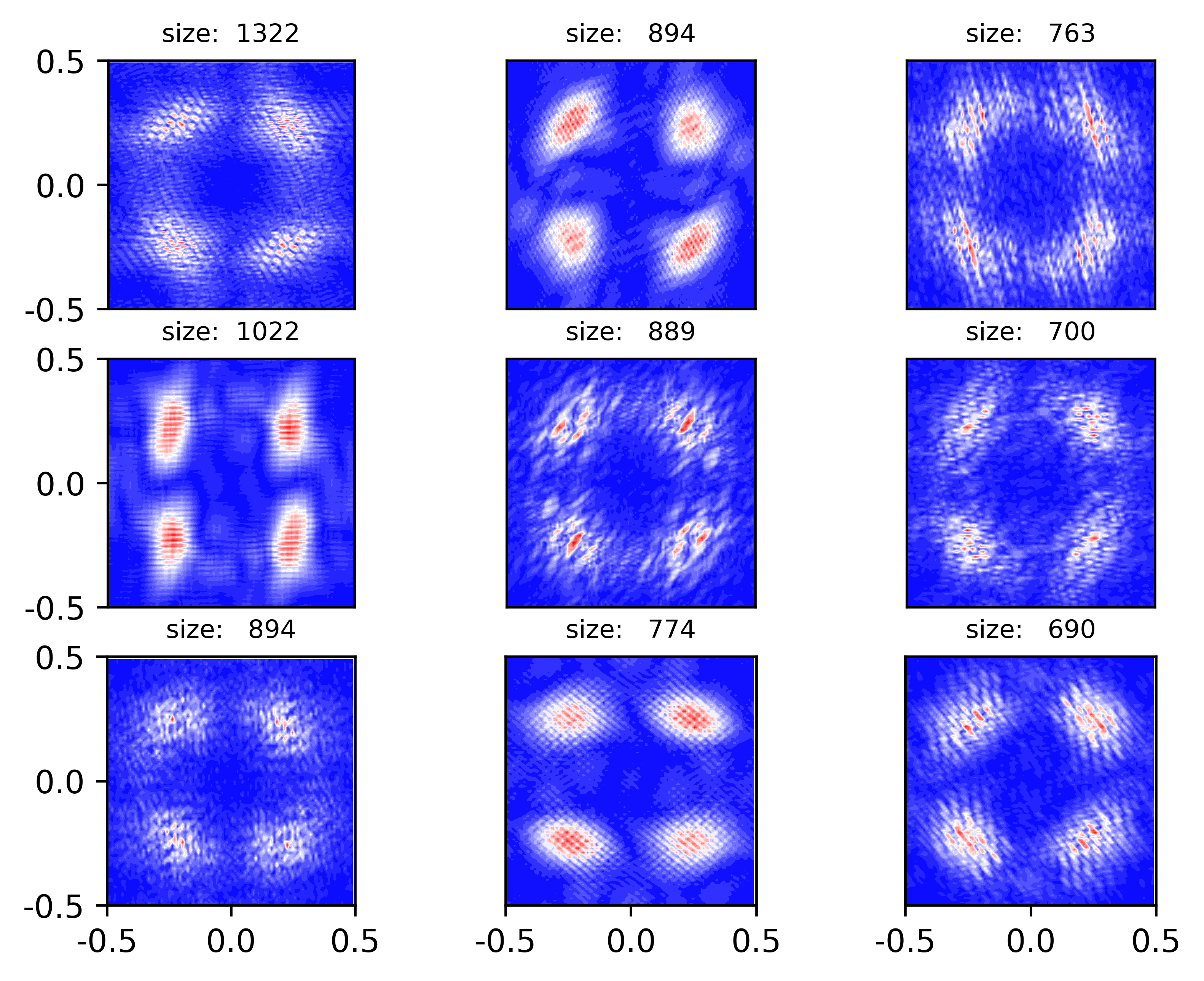}
\caption{Fourier transforms of individual clusters, with sizes in number of Fe atoms above each panel. Units of $2\pi/a$.}
\label{figsingle_clusters}
\end{figure*}
The eigenvalue problem of this huge, but sparse matrix was solved using the \emph{python-scipy} package. The obtained eigenvector was Fourier-transformed using the FFT-module in the \emph{python-scipy} package. The results are illustrated in Fig.~\ref{figsimulations}f in the main text. Fourier transforms of some larger individual clusters are given in Fig.~\ref{figsingle_clusters}, showing that the magnetism of the whole sample is an amplification of the cluster responses.

\begin{acknowledgments}
The work at the University of Zagreb was supported by project CeNIKS co-financed by the Croatian Government and the European Union through the European Regional Development Fund -Competitiveness and Cohesion Operational Program (Grant No. KK.01.1.1.02.0013), the Croatian-Swiss Research Program of the Croatian Science Foundation and the Swiss National Science Foundation with funds obtained from the Swiss-Croatian Cooperation Program Project No. IZHRZ0-180652, the Swiss National Science Foundation through grant No. 200021-175836, and by the Croatian Science Foundation under Project No. IP-2022-10-3382.
The work at AGH University was supported by the National Science Centre, Poland, Grant OPUS: 2021/41/B/ST3/03454; the Polish National Agency for Academic Exchange under “Polish Returns 2019” Programme No. PPN/PPO/2019/1/00014; and the “Excellence Initiative-Research University” program for AGH University of Krakow. I.B. acknowledges support from the Swiss Confederation through Government Excellence Scholarships.
L.A. acknowledges the support of Austrian Science Fund (FWF) F86.
J.R acknowledges support from the Singapore National Science Scholarship, Agency for Science Technology and Research and the European Research Council (HERO, Grant No. 810451).
We acknowledge Virgile Favre (ILL, D20, Proposal ID: 5-31-2772), Bachir Ouladdiaf (ILL, Cyclops, Proposal ID: EASY-1015) and Paola Caterina (PSI, Zebra, Proposal ID: 20220994) for their help with proposal preparation and acquisition of the neutron diffraction data. 
The work at TU Wien was supported by FWF Project P 35945-N.
\end{acknowledgments}

\bibliographystyle{unsrtnat}
\bibliography{High-entropy_magnetism_of_murunskite}

\begin{thebibliography}{31}
\providecommand{\natexlab}[1]{#1}
\providecommand{\url}[1]{\texttt{#1}}
\expandafter\ifx\csname urlstyle\endcsname\relax
  \providecommand{\doi}[1]{doi: #1}\else
  \providecommand{\doi}{doi: \begingroup \urlstyle{rm}\Url}\fi

\bibitem[Schr\"odinger(1992)]{Schrodinger92}
Erwin Schr\"odinger.
\newblock \emph{What is life? : the physical aspect of the living cell ; with Mind and matter ; {\&} Autobiographical sketches}.
\newblock Canto. Cambridge University Press Cambridge, Cambridge, 1992.
\newblock ISBN 0521427088; 9780521427081; 9781107604667; 1107604664.

\bibitem[Anderson(1972)]{Anderson72}
P.~W. Anderson.
\newblock More is different.
\newblock \emph{Science}, 177\penalty0 (4047):\penalty0 393--396, 1972.
\newblock \doi{10.1126/science.177.4047.393}.

\bibitem[Shen and Davis(2008)]{Shen08}
Kyle~M. Shen and J.C.~Seamus Davis.
\newblock Cuprate high-t$_c$ superconductors.
\newblock \emph{Materials Today}, 11\penalty0 (9):\penalty0 14--21, 2008.
\newblock ISSN 1369-7021.
\newblock \doi{10.1016/S1369-7021(08)70175-5}.

\bibitem[Yoshida et~al.(2012)Yoshida, Hashimoto, M.~Vishik, Shen, and Fujimori]{Yoshida12}
Teppei Yoshida, Makoto Hashimoto, Inna M.~Vishik, Zhi-Xun Shen, and Atsushi Fujimori.
\newblock Pseudogap, superconducting gap, and fermi arc in high-tc cuprates revealed by angle-resolved photoemission spectroscopy.
\newblock \emph{Journal of the Physical Society of Japan}, 81\penalty0 (1):\penalty0 011006, 2012.
\newblock \doi{10.1143/JPSJ.81.011006}.

\bibitem[Si et~al.(2016)Si, Yu, and Abrahams]{Si16}
Qimiao Si, Rong Yu, and Elihu Abrahams.
\newblock High-temperature superconductivity in iron pnictides and chalcogenides.
\newblock \emph{Nature Reviews Materials}, 1\penalty0 (4):\penalty0 16017, Mar 2016.
\newblock ISSN 2058-8437.
\newblock \doi{10.1038/natrevmats.2016.17}.

\bibitem[Tolj et~al.(2021)Tolj, {Iv\v si\'c}, {\v Zivkovi\'c}, Semeniuk, Martino, Akrap, Reddy, Klebel-Knobloch, {Lon\v cari\'c}, {Forr\'o}, {Bari\v si\'c}, Ronnow, and Sunko]{Tolj21}
Davor Tolj, Trpimir {Iv\v si\'c}, Ivica {\v Zivkovi\'c}, Konstantin Semeniuk, Edoardo Martino, Ana Akrap, Priyanka Reddy, Benjamin Klebel-Knobloch, Ivor {Lon\v cari\'c}, {L\'aszl\'o} {Forr\'o}, Neven {Bari\v si\'c}, Henrik~M. Ronnow, and Denis~K. Sunko.
\newblock Synthesis of murunskite single crystals: A bridge between cuprates and pnictides.
\newblock \emph{Applied Materials Today}, 24:\penalty0 101096, 2021.
\newblock \doi{10.1016/j.apmt.2021.101096}.

\bibitem[Zaanen et~al.(1985)Zaanen, Sawatzky, and Allen]{Zaanen85}
J.~Zaanen, G.~A. Sawatzky, and J.~W. Allen.
\newblock Band gaps and electronic structure of transition-metal compounds.
\newblock \emph{Phys. Rev. Lett.}, 55:\penalty0 418--421, Jul 1985.
\newblock \doi{10.1103/PhysRevLett.55.418}.

\bibitem[Eskes and Jefferson(1993)]{Eskes93}
Henk Eskes and John~H. Jefferson.
\newblock Superexchange in the cuprates.
\newblock \emph{Phys. Rev. B}, 48:\penalty0 9788--9798, Oct 1993.
\newblock \doi{10.1103/PhysRevB.48.9788}.

\bibitem[Barišić et~al.(2015)Barišić, Chan, Veit, Dorow, Ge, Tang, Tabis, Yu, Zhao, and Greven]{Barisic15}
N.~Barišić, M.~K. Chan, M.~J. Veit, C.~J. Dorow, Y.~Ge, Y.~Tang, W.~Tabis, G.~Yu, X.~Zhao, and M.~Greven.
\newblock Evidence for a universal {Fermi}-liquid scattering rate throughout the phase diagram of the copper-oxide superconductors.
\newblock \emph{arXiv:1507.07885 [cond-mat]}, July 2015.
\newblock URL \url{http://arxiv.org/abs/1507.07885}.
\newblock arXiv: 1507.07885 version: 1.

\bibitem[Pelc et~al.(2019)Pelc, Popčević, Požek, Greven, and Barišić]{Pelc19}
D.~Pelc, P.~Popčević, M.~Požek, M.~Greven, and N.~Barišić.
\newblock Unusual behavior of cuprates explained by heterogeneous charge localization.
\newblock \emph{Science Advances}, 5\penalty0 (1):\penalty0 eaau4538, January 2019.
\newblock \doi{10.1126/sciadv.aau4538}.

\bibitem[Barišić and Sunko(2022)]{Barisic22}
N.~Barišić and D.~K. Sunko.
\newblock High-{T}$_{\textrm{c}}$ {Cuprates}: a story of two electronic subsystems.
\newblock \emph{Journal of Superconductivity and Novel Magnetism}, 35\penalty0 (7):\penalty0 1781--1799, July 2022.
\newblock \doi{10.1007/s10948-022-06183-y}.

\bibitem[Kumar et~al.(2023)Kumar, Akrap, Homes, Martino, Klebel-Knobloch, Tabis, Bari\ifmmode \check{s}\else \v{s}\fi{}i\ifmmode~\acute{c}\else \'{c}\fi{}, Sunko, and Bari\ifmmode \check{s}\else \v{s}\fi{}i\ifmmode~\acute{c}\else \'{c}\fi{}]{Kumar23}
C.~M.~N. Kumar, A.~Akrap, C.~C. Homes, E.~Martino, B.~Klebel-Knobloch, W.~Tabis, O.~S. Bari\ifmmode \check{s}\else \v{s}\fi{}i\ifmmode~\acute{c}\else \'{c}\fi{}, D.~K. Sunko, and N.~Bari\ifmmode \check{s}\else \v{s}\fi{}i\ifmmode~\acute{c}\else \'{c}\fi{}.
\newblock Characterization of two electronic subsystems in cuprates through optical conductivity.
\newblock \emph{Phys. Rev. B}, 107:\penalty0 144515, Apr 2023.
\newblock \doi{10.1103/PhysRevB.107.144515}.

\bibitem[Bari\ifmmode \check{s}\else \v{s}\fi{}i\ifmmode~\acute{c}\else \'{c}\fi{} et~al.(2010)Bari\ifmmode \check{s}\else \v{s}\fi{}i\ifmmode~\acute{c}\else \'{c}\fi{}, Wu, Dressel, Li, Cao, and Xu]{Barisic10}
N.~Bari\ifmmode \check{s}\else \v{s}\fi{}i\ifmmode~\acute{c}\else \'{c}\fi{}, D.~Wu, M.~Dressel, L.~J. Li, G.~H. Cao, and Z.~A. Xu.
\newblock Electrodynamics of electron-doped iron pnictide superconductors: Normal-state properties.
\newblock \emph{Phys. Rev. B}, 82:\penalty0 054518, Aug 2010.
\newblock \doi{10.1103/PhysRevB.82.054518}.

\bibitem[Eschrig and Koepernik(2009)]{Eschrig09}
Helmut Eschrig and Klaus Koepernik.
\newblock Tight-binding models for the iron-based superconductors.
\newblock \emph{Phys. Rev. B}, 80:\penalty0 104503, Sep 2009.
\newblock \doi{10.1103/PhysRevB.80.104503}.

\bibitem[Fink et~al.(2009)Fink, Thirupathaiah, Ovsyannikov, D\"urr, Follath, Huang, de~Jong, Golden, Zhang, Jeschke, Valent\'{\i}, Felser, Dastjani~Farahani, Rotter, and Johrendt]{Fink09}
J.~Fink, S.~Thirupathaiah, R.~Ovsyannikov, H.~A. D\"urr, R.~Follath, Y.~Huang, S.~de~Jong, M.~S. Golden, Yu-Zhong Zhang, H.~O. Jeschke, R.~Valent\'{\i}, C.~Felser, S.~Dastjani~Farahani, M.~Rotter, and D.~Johrendt.
\newblock Electronic structure studies of ${\text{bafe}}_{2}{\text{as}}_{2}$ by angle-resolved photoemission spectroscopy.
\newblock \emph{Phys. Rev. B}, 79:\penalty0 155118, Apr 2009.
\newblock \doi{10.1103/PhysRevB.79.155118}.

\bibitem[Borisenko et~al.(2016)Borisenko, Evtushinsky, Liu, Morozov, Kappenberger, Wurmehl, Buchner, Yaresko, Kim, Hoesch, Wolf, and Zhigadlo]{Borisenko16}
S.~V. Borisenko, D.~V. Evtushinsky, Z.-H. Liu, I.~Morozov, R.~Kappenberger, S.~Wurmehl, B.~Buchner, A.~N. Yaresko, T.~K. Kim, M.~Hoesch, T.~Wolf, and N.~D. Zhigadlo.
\newblock Direct observation of spin-orbit coupling in iron-based superconductors.
\newblock \emph{Nat Phys}, 12\penalty0 (4):\penalty0 311--317, Apr 2016.
\newblock ISSN 1745-2473.
\newblock \doi{10.1038/nphys3594}.
\newblock Letter.

\bibitem[Derondeau et~al.(2017)Derondeau, Bisti, Kobayashi, Braun, Ebert, Rogalev, Shi, Schmitt, Ma, Ding, Strocov, and Min\'ar]{Derondeau17}
Gerald Derondeau, Federico Bisti, Masaki Kobayashi, J\"urgen Braun, Hubert Ebert, Victor~A. Rogalev, Ming Shi, Thorsten Schmitt, Junzhang Ma, Hong Ding, Vladimir~N. Strocov, and J\'an Min\'ar.
\newblock Fermi surface and effective masses in photoemission response of the {(Ba$_{1-x}$K$_x$)Fe$_2$As$_2$} superconductor.
\newblock \emph{Scientific Reports}, 7\penalty0 (1):\penalty0 8787, 2017.
\newblock \doi{10.1038/s41598-017-09480-y}.

\bibitem[Sunko(2020)]{Sunko20a}
D.~K. Sunko.
\newblock High-temperature superconductors as ionic metals.
\newblock \emph{Journal of Superconductivity and Novel Magnetism}, 33\penalty0 (1):\penalty0 27--33, Jan 2020.
\newblock \doi{10.1007/s10948-019-05280-9}.

\bibitem[Moniri et~al.(2023)Moniri, Yang, Ding, Yuan, Zhou, Yang, Zhu, Liao, Yao, Hu, Ercius, and Miao]{Moniri23}
Saman Moniri, Yao Yang, Jun Ding, Yakun Yuan, Jihan Zhou, Long Yang, Fan Zhu, Yuxuan Liao, Yonggang Yao, Liangbing Hu, Peter Ercius, and Jianwei Miao.
\newblock Three-dimensional atomic structure and local chemical order of medium- and high-entropy nanoalloys.
\newblock \emph{Nature}, 624\penalty0 (7992):\penalty0 564--569, Dec 2023.
\newblock ISSN 1476-4687.
\newblock \doi{10.1038/s41586-023-06785-z}.

\bibitem[Campbell et~al.(2006)Campbell, Stokes, Tanner, and Hatch]{Campbell06}
Branton~J. Campbell, Harold~T. Stokes, David~E. Tanner, and Dorian~M. Hatch.
\newblock {{\it ISODISPLACE}: a web-based tool for exploring structural distortions}.
\newblock \emph{Journal of Applied Crystallography}, 39\penalty0 (4):\penalty0 607--614, Aug 2006.
\newblock \doi{10.1107/S0021889806014075}.
\newblock URL \url{https://doi.org/10.1107/S0021889806014075}.

\bibitem[Aroyo et~al.(2011)Aroyo, Perez-Mato, Orobengoa, Tasci, de~la Flor, and Kirov]{Aroyo11}
Mois~I Aroyo, Juan~Manuel Perez-Mato, Danel Orobengoa, EMRE Tasci, Gemma de~la Flor, and Asel Kirov.
\newblock Crystallography online: Bilbao crystallographic server.
\newblock \emph{Bulg. Chem. Commun}, 43\penalty0 (2):\penalty0 183--197, 2011.

\bibitem[Perez-Mato et~al.(2015)Perez-Mato, Gallego, Tasci, Elcoro, de~la Flor, and Aroyo]{Perez15}
JM~Perez-Mato, SV~Gallego, ES~Tasci, LU{\.I}S Elcoro, Gemma de~la Flor, and MI~Aroyo.
\newblock Symmetry-based computational tools for magnetic crystallography.
\newblock \emph{Annual Review of Materials Research}, 45:\penalty0 217--248, 2015.

\bibitem[Qureshi(2019)]{Qureshi19}
Navid Qureshi.
\newblock {{\it Mag2Pol}: a program for the analysis of spherical neutron polarimetry, flipping ratio and integrated intensity data}.
\newblock \emph{Journal of Applied Crystallography}, 52\penalty0 (1):\penalty0 175--185, Feb 2019.
\newblock \doi{10.1107/S1600576718016084}.

\bibitem[Duraj and Ruebenbauer(2011)]{Duraj11}
\L. Duraj and K.~Ruebenbauer.
\newblock {New Graphical Interface to the MOSGRAF Suite}.
\newblock \emph{Acta Physica Polonica A}, 119\penalty0 (1):\penalty0 75--77, 2011.
\newblock \doi{10.12693/APhysPolA.119.75}.

\bibitem[Blundell(2001)]{Blundell01}
Stephen Blundell.
\newblock \emph{{Magnetism in Condensed Matter}}.
\newblock Oxford Master Series in Condensed Matter Physics. Oxford University Press, Oxford, 2001.
\newblock ISBN 9780198505914.

\bibitem[Lei et~al.(2011)Lei, Abeykoon, Bozin, and Petrovic]{Lei11-180503}
Hechang Lei, Milinda Abeykoon, Emil~S. Bozin, and C.~Petrovic.
\newblock Spin-glass behavior of semiconducting ${\mathrm{k}}_{x}$fe${}_{2\ensuremath{-}y}$s${}_{2}$.
\newblock \emph{Phys. Rev. B}, 83:\penalty0 180503, May 2011.
\newblock \doi{10.1103/PhysRevB.83.180503}.

\bibitem[Popescu and Zunger(2012)]{Popescu12}
Voicu Popescu and Alex Zunger.
\newblock Extracting {$E$} versus $\vec{k}$ effective band structure from supercell calculations on alloys and impurities.
\newblock \emph{Phys. Rev. B}, 85:\penalty0 085201, Feb 2012.
\newblock \doi{10.1103/PhysRevB.85.085201}.

\bibitem[Zhao et~al.(2009)Zhao, Adroja, Yao, Bewley, Li, Wang, Wu, Chen, Hu, and Dai]{Zhao09}
J.~Zhao, D.~T. Adroja, D.~Yao, R.~Bewley, S.~Li, X.~F. Wang, G.~Wu, X.~H. Chen, J.~Hu, and P.~Dai.
\newblock Spin waves and magnetic exchange interactions in cafe2as2.
\newblock \emph{Nature Physics}, 5:\penalty0 555–560, Jul 2009.
\newblock \doi{10.1038/nphys1336}.

\bibitem[Klebel-Knobloch et~al.(2023)Klebel-Knobloch, Tabi{\'{s}}, Gala, Bari{\v{s}}i{\'{c}}, Sunko, and Bari{\v{s}}i{\'{c}}]{Klebel23}
B.~Klebel-Knobloch, W.~Tabi{\'{s}}, M.~A. Gala, O.~S. Bari{\v{s}}i{\'{c}}, D.~K. Sunko, and N.~Bari{\v{s}}i{\'{c}}.
\newblock Transport properties and doping evolution of the fermi surface in cuprates.
\newblock \emph{Scientific Reports}, 13\penalty0 (1):\penalty0 13562, Aug 2023.
\newblock ISSN 2045-2322.
\newblock \doi{10.1038/s41598-023-39813-z}.

\bibitem[Palle and Sunko(2021)]{Palle21}
Grgur Palle and D~K Sunko.
\newblock Physical limitations of the hohenberg–mermin–wagner theorem.
\newblock \emph{Journal of Physics A: Mathematical and Theoretical}, 54\penalty0 (31):\penalty0 315001, jul 2021.
\newblock \doi{10.1088/1751-8121/ac0a9d}.

\bibitem[Gütlich et~al.(21)Gütlich, Schröder, and Schünemann]{Gütlich2012}
P.~Gütlich, C.~Schröder, and V.~Schünemann.
\newblock Mössbauer spectroscopy—an indispensable tool in solid state research.
\newblock \emph{Spectroscopy Europe}, 24\penalty0 (4), 21.

\end{thebibliography}

\end{document}